# Fewer Qubits, Better Choices: Coupling-Aware Sub-QUBO Selection for Quantum-Assisted Traffic Zone Partitioning

Qianwen (Vivian) Guo, Ph.D.
Department of Civil & Environmental Engineering
FAMU-FSU College of Engineering
1753 W Paul Dirac Dr, Tallahassee, FL 32310
Phone: +1(850) 410-6252;
E-mail: qguo@eng.famu.fsu.edu

Ruimin Ke, Ph.D. (Corresponding Author)
Department of Civil & Environmental Engineering
Rensselaer Polytechnic Institute
110 8th Street, Troy, NY 12180
Phone: +1 (518)276-7097;
E-mail: ker@rpi.edu

## Abstract

Near-term quantum optimizers accommodate roughly a hundred binary variables, requiring large Quadratic Unconstrained Binary Optimization problems to be decomposed into hardware-sized subproblems while other variables remain fixed. Variables are typically selected by ranking the objective changes from individual flips. We show that this rule is uninformative once the incumbent is optimal under single-variable flips: every score represents a cost, while remaining improvements depend on overlooked pairwise interactions. A second-order expansion instead yields a prize-collecting densest-subgraph selection problem, solvable greedily in time linear in the number of variables, and two solver-free bounds bracketing the improvement achievable within a selected subset. On transportation networks from Chicago and Philadelphia, our rule with 16-variable subproblems outperforms random selection with 64, demonstrating that quadrupling device capacity cannot compensate for weaker selection. Replacing geometric adjacency with road-network connectivity widens the margin by a factor of 1.6–1.9. With the selection trace fixed, varying the fraction of subproblems solved on a 120-qubit device from zero to one leaves the final objective unchanged to six significant figures. At tested hardware-accessible sizes, gains therefore arise from selection rather than the backend. Hardware feasibility is limited by coupling terms: 120-variable problems fail to compile with 7,260 terms but succeed with 5,118. Compilation scales almost quadratically with term count and dominates execution cost, reaching 1,752 seconds versus 469 seconds of quantum processing.

# 1 Introduction

Dividing a transportation network into a small number of internally coherent traffic analysis zones is a prerequisite for almost every downstream planning and control task. Demand modeling, network-level signal and perimeter control, hierarchical routing, and macroscopic traffic-state estimation all operate on zones rather than on individual links or intersections (Ortúzar & Willumsen, 2024). How that partition is drawn is not a cosmetic choice. Ji and Geroliminis (2012) showed that partitioning a heterogeneous network into spatially compact, internally homogeneous subregions is a precondition for a well-defined network-level relationship between accumulation and flow, and that a poorly drawn partition degrades the reliability of any control scheme built on top of it. This principle has been foundational in subsequent network partitioning research (Geroliminis & Daganzo, 2008; Geroliminis & Sun, 2011). The same partitioning step underlies macroscopic-fundamental-diagram-based hierarchical network management (An et al., 2018), multimodal network partitioning (Saeedmanesh & Geroliminis, 2016; Geroliminis et al, 2014), and the increasingly common practice of combining topological connectivity with observed flow attributes when defining zone boundaries (Ma et al., 2023). Because the resulting subregions are used for real operational decisions downstream, the boundary itself has to minimize the flow that is cut between zones while keeping each zone spatially coherent and reasonably balanced in size (McNally, 2007).

The specific problem of drawing zone boundaries to minimize inter-zone cut, subject to spatial-coherence and balance requirements, has accordingly attracted a substantial and still-growing literature. Optimal-transport and distance-based clustering methods have been used to jointly exploit network topology and observed traffic attributes when recovering coherent subregions (Gavra et al., 2025), and dedicated work has extended the basic partitioning objective to weigh network connectivity against traffic attributes rather than geometry alone (Ma et al., 2023). Graph-contrastive and learning-based partitioning frameworks have been proposed to make the resulting zones adapt to changing demand (Hu et al., 2025). Community-detection methods developed outside transportation, including normalized-cut spectral partitioning (Shi & Malik, 2000) and modularity-based community detection (Fortunato, 2010), supply much of the algorithmic machinery that zone-partitioning studies build on. This body of work establishes that the underlying decision problem, partitioning a weighted graph so as to minimize the cut between parts subject to side constraints, is not merely difficult in practice but is NP-hard in general (Garey & Johnson, 1979), and the number of candidate partitions grows combinatorially with the number of zones. Even the restricted densest-k-subgraph problem that this paper's selector reduces to admits no known polynomial-time approximation scheme (Khot, 2006). As real transportation networks are discretized into hundreds or thousands of traffic analysis zones, exact classical solution of the resulting partitioning problem is out of reach, and even strong classical heuristics scale poorly once the zone count reaches the sizes seen in Chicago-Sketch or Philadelphia.

Quantum optimization has been proposed as one route around this scaling wall. Because the partitioning objective is naturally quadratic in binary zone-assignment variables, it maps directly onto the Ising/ Quadratic Unconstrained Binary Optimization (QUBO) form that near-term quantum annealers and digitized counterdiabatic optimizers are built to accept, and combinatorial problems of exactly this shape are the setting in which quantum devices are argued to offer the clearest near-term advantage over classical heuristics (Farhi et al., 2014; Rønnow et al., 2014). Quantum annealing has already been applied to graph partitioning directly (Ushijima-Mwesigwa et al., 2017) and to the closely related traffic-flow optimization problem (Neukart et al., 2017), and a broader survey of quantum methods across routing, dispatching and network restoration in transportation confirms that the field has converged on the QUBO model as its common input format (Azfar & Ke, 2026a; Zhuang et al., 2024; Udekwe et al., 2026; Azfar et al., 2025). That convergence is not specific to transportation: the Quadratic Unconstrained Binary Optimization model has become, in the words of its most-cited tutorial, the most widely applied optimization model in the quantum computing area, unifying scheduling, routing, portfolio, max-cut and partitioning problems within a single formulation that both quantum annealers and gate-based optimizers accept as input (Glover et al., 2019), and a dedicated survey catalogs the resulting breadth of unconstrained binary quadratic programming applications across operations research (Kochenberger et al., 2014). The Ising-formulation catalog of Lucas (2014) performs the same unifying role for the physics side of the same model.

The obstacle is size rather than expressiveness. Near-term devices accept on the order of a hundred programmable binary variables, while the zone-partitioning instances practitioners care about have thousands, and the mismatch is structural rather than a matter of waiting for larger processors: as this paper's own hardware experiments, the binding hardware limit is not even the qubit count but the number of pairwise coupling terms the compiler must route, so simply adding qubits does not remove the constraint. Dense zone-adjacency structure at real city scale routinely produces QUBO instances with thousands of quadratic terms, well beyond what a single device call can compile, let alone solve. The standard response, adopted by the reference decomposition solver qbsolv (Booth et al., 2017) and used across the sub-QUBO literature (Atobe et al., 2022; Zhao & Tang, 2025; Bass et al., 2021; Rosenberg et al., 2016; Okada et al., 2019; Shaydulin et al., 2019), is to decompose the large QUBO into a sequence of hardware-sized sub-problems: hold most zone-assignment variables fixed, release a small subset of size q for the device to solve, write the result back, and repeat. This sub-QUBO strategy is what allows a partitioning problem with thousands of zones to be attacked on a device with only a hundred or so usable variables, without waiting for the qubit count itself to grow. We adopted exactly this decomposition for traffic zone partitioning in our own earlier work (Ke et al., 2026), where sub-problems were selected by ranking variables on their individual objective impact, the rule this paper revisits and replaces. The outer loop spends nearly all of its time at solutions that are already optimal under single-variable flips, and at such a point every single-flip score is a pure cost. All remaining improvement lies in the interactions between variables, and the rule never examines them.

This suggests replacing the per-variable score with an objective computed from the interactions themselves. Expanding the QUBO to second order about the incumbent yields two quantities: a vector a of individual flip costs, and a matrix $K$ of pairwise corrections. Choosing a subset then becomes a prize-collecting densest-subgraph problem, find $q$ variables that are strongly coupled to one another while individually cheap, which we solve greedily in time linear in the number of variables. The same expansion supplies two bounds, computable in time quadratic in the sub-problem size and without any solver call, that bracket the improvement a chosen subset can deliver.

We evaluate this on real transportation networks from Chicago, IL and Philadelphia, PA from the Transportation Networks for Research repository. The contributions of this study are threefold.

We demonstrate that structure-aware variable selection can substitute for increased hardware capacity. On the 1,525-zone Philadelphia, PA network, our coupling-aware selection rule with a sub-problem size of only 16 variables outperforms random selection using 64 variables. This result shows that a fourfold increase in device capacity cannot compensate for a weak selection strategy. More importantly, this finding does not depend on quantum hardware; rather, it demonstrates that selecting the right variables can be more important than simply increasing the size of the sub-problem that the hardware can accommodate.

We show that incorporating transportation-domain structure substantially strengthens the effectiveness of coupling-aware selection. By replacing conventional geometric adjacency between zone centroids with connectivity derived from the actual road network, we change approximately 62% of the adjacency edges in the larger instance and increase the advantage of the proposed selection rule by roughly 1.6-1.9 times. The road-network representation yields a sparser and more discriminative interaction structure, enabling our method to identify influential variables more effectively. This result also helps explain the behavior of our early synthetic benchmark, in which a dominant balance term produced an almost uniform coupling structure and therefore left little opportunity for any structure-aware rule to outperform random selection.

We establish that the observed gains are attributable to the classical selection strategy rather than the quantum sub-solver, while also identifying the main practical hardware bottleneck. When we vary the fraction of selected sub-problems sent to quantum hardware from 0% to 100% while holding the selection trace fixed, the final objective remains identical across all five settings to every reported digit. We therefore position our method as a classical structure-aware selection framework that can use a quantum processor as a backend, rather than as a demonstration of quantum advantage. Our hardware experiments further show that scalability is constrained more by interaction density than by qubit count: a full-width 120-variable problem fails under dense coupling but succeeds after the coupling terms are thinned.

This paper is organized as follows. Section 2 places the work against sub-QUBO decomposition, working-set selection in convex optimization, densest-subgraph problems, and quantum benchmarking practice. Section 3 develops the flip-space expansion, the selection objective and its bounds. Section 4 describes the instances and protocol, Section 5 reports the experiments, Section 6 collects the transferable practical conclusions, and Section 7 concludes.

## 2 Literature review

### 2.1 Sub-QUBO decomposition and its selection rules

Splitting a QUBO too large for the available solver into hardware-sized sub-problems, solving each with the remaining variables clamped, and iterating, is an established pattern. The reference implementation is qbsolv (Booth et al., 2017) whose selection rule sorts variables by impact, the increase in objective value when a variable is negated at the incumbent, and cuts contiguous blocks out of that ordered list. The same rule was carried into traffic zone partitioning by (Ke et al., 2026), which is the direct predecessor of the present work. In the notation of Section 3.3, impact is $|a_i|$, so both (Booth et al., 2017) and (Ke et al., 2026) use precisely the rule analyzed in Section 3.5, and its blindness to K at a 1-opt optimum is a property of the most widely used decomposition solver, and of our own prior method, rather than of a straw man.

Two other rules appear in the literature. Atobe et al. (2022) maintain a pool of solution instances and extract the variables of least certainty, measured by how evenly a variable is split across the pool. That rule reads disagreement between solutions rather than problem structure, and it requires a population of solutions to be maintained. Zhao and Tang (2025) build a variable-correlation matrix and cluster it, grouping correlated variables into a common sub-problem. Theirs is the closest prior work to ours in spirit, because it also reads structure rather than a per-variable score, and it likewise compares against Booth et al. (2017) and Atobe et al. (2022). It differs in what the structure is derived from and in what it delivers: the correlation matrix is an empirical object estimated over solutions, whereas the objective in Section 3.6 is derived in closed form from the exact second-order expansion of the QUBO at the incumbent, and comes with bounds (Section 3.7) that a clustering criterion does not provide.

Beyond the choice of rule, decomposition has been studied empirically and structurally. Bass et al. (2021) benchmark decomposition strategies and find that propagating each solved sub-problem back into the remainder is what matters most, the field-folding step of Section 3.2. Rosenberg et al. (2016) build an iterative sub-problem solver around an annealer. Okada et al. (2019) choose sub-problems so as to maximize what embeds on the hardware graph, which makes the selection criterion a property of the device rather than of the problem. Shaydulin et al. (2019) decompose by community structure of the problem graph. None of these scores candidate subsets by the improvement they could yield.

### 2.2 Working-set selection outside quantum computing

The same problem, choose a small subset of variables to re-optimize while the rest stay fixed, is the core of decomposition methods for support vector machines, and that literature has already been through the transition we are proposing. Platt's (1998) SMO takes the extreme case of two variables per sub-problem, chosen by KKT-violation heuristics. Joachims's (1999) SVM-light generalizes to a fixed working set of size $q$ selected by Zoutendijk's steepest feasible direction: the $q$ variables with the best first-order rate of descent. That criterion is, structurally, impact indexing, a per-variable first-order score, blind to interactions within the selected set.

Fan et al. (2005) then showed that selecting the working set by the actual second-order decrease of the objective, rather than by the first-order rate, gives materially faster convergence. The move from $a$ alone to $a$ together with $K$ in Section 3.6 is the same move, transposed from a convex QP to a binary quadratic problem. We regard this as the strongest argument that the idea is sound rather than merely convenient: it is a known improvement in a mature literature that the sub-QUBO line of work has not yet adopted.

Large-neighborhood search (Shaw, 1998; Ropke & Pisinger, 2006) and variable-neighborhood search (Mladenović & Hansen, 1997) also destroy and repair part of a solution, and are the honest ancestors of the outer loop. They differ in what governs the choice. Shaw (1998) removes customers by a hand-designed relatedness measure. Adaptive LNS (Ropke & Pisinger, 2006) learns weights over a portfolio of destroy operators from past success, and VNS (Mladenović & Hansen, 1997) escalates through a pre-declared family of neighborhoods. All three treat the neighborhood as a design input. What we add is a neighborhood chosen by an objective derived from the problem instance itself, with a computable bracket on what that choice is worth.

### 2.3 Densest k-subgraph

The selector of Section 3.6 is an instance of a well-studied combinatorial problem. Densest k-subgraph is NP-hard, contains max-clique as a special case, and admits no PTAS under the assumption that NP has no randomized subexponential-time algorithms (Khot, 2006). The best known approximation is $O(n^{1/4+\varepsilon})$ (Bhaskara et al., 2010), improving the earlier $O(n^{\delta})$ with $\delta < \frac{1}{3}$ of Feige et al. (2001). Khuller and Saha (2009) give 2-approximations for the at-least-k variant and relate the at-most-k variant back to DkS. Greedy peeling, the natural heuristic, has a tight ratio of about $2n/k$ (Asahiro et al., 2000), good only when $k$ is a constant fraction of $n$, which is emphatically not our regime, where $q \ll N$.

We therefore make no approximation claim for the selector. This is a deliberate consequence of the theory rather than an omission: the greedy guarantee of Nemhauser et al. (1978) requires monotone submodularity, and $F(S)$ is monotone but not submodular (in Section 3.4), while the DkS guarantees that do apply are too weak at $q \ll N$ to be worth quoting. What we offer instead is the per-round certificate of Section 3.7, which brackets the value of the subset actually selected rather than bounding the algorithm's worst case.

Our objective subtracts a per-node cost $a_i$ from the edge density. We could not find this variant in the literature: the standard survey of densest-subgraph variants (Tsourakakis et al., 2013) catalogs size penalties and additive node weights, but not subtractive node costs. We state it as a formulation of convenience derived from the QUBO expansion, not as a claim of a new graph-theoretic problem.

### 2.4 Quantum optimization and benchmarking practice

The sub-solver used here is bias-field digitized counterdiabatic quantum optimization (Gomez Cadavid et al., 2025), in the higher-order form (Romero et al., 2025) that the Iskay service implements, with hardware benchmarks reported in (Chandarana et al., 2025). A parallel line of work attacks the same hardware constraints from the circuit side rather than by decomposition, through compressed adiabatic evolution (Azfar et al., 2026) and through shallower, more robust variational schedules (Azfar & Ke, 2026 b). Those techniques are complementary to selection and could be combined with it. The hardware benchmarks claim runtime advantages against classical baselines at problem sizes and structures different from ours. Our Section 5.6 finding that the backend is substitutable at $q = 16$ does not contradict them, because at that size our classical sub-solver is exact and no backend can do better. The two results address different regimes and we are careful not to conflate them.

How such comparisons should be reported is itself a contested question. Rønnow and colleagues (Rønnow et al., 2014) set out what does and does not constitute evidence of speedup, and their argument that the comparison must be made against a well-tuned classical baseline on a matched metric is why Section 5.7 reports QPU time from provider usage records rather than wall-clock time. Enan and colleagues (Enan et al., 2024) document the practical consequence in a transportation setting: their end-to-end latency is 2.99 s of which only 0.197 s is QPU access, a 6.6% share, against the 5.7% we measure at a problem scale two orders of magnitude larger. Queueing, not computation, sets wall-clock time on current cloud services.

### 2.5 Partitioning in transportation

Mapping combinatorial problems to Ising or QUBO form is cataloged by Lucas (2014), and graph partitioning has been run on annealing hardware by Ushijima-Mwesigwa et al. (2017). The best-known transportation application of quantum annealing is the traffic-flow optimization of Neukart et al. (2017). The broader landscape of quantum methods in intelligent transportation systems is surveyed in Zhuang et al. (2024), and gate-model applications now include vehicle routing (Azfar et al., 2025), cooperative platoon routing and dispatching (Azfar & Ke, 2026), and resilient network restoration (Udekwe et al., 2026). On the classical side, spatial partitioning of urban transportation networks is a mature topic, Ji and Geroliminis (2012) is the standard reference, and recent work combines network connectivity with flow attributes in the way our road-network adjacency does (Ma et al., 2023). Our contribution to this literature is narrow and specific: Section 5.4 shows that replacing geometric adjacency with road-network reachability does not merely change the instance, it widens the advantage of a coupling-aware selection rule, because the sparser adjacency makes the coupling structure more discriminative

### 2.6 Summary

Selecting a subset of variables to re-optimize is not novel, and we do not claim it. What this paper contributes is a selection objective derived in closed form from the second-order expansion of the QUBO at the incumbent, rather than from a per-variable first-order score or an empirical correlation estimate. It contributes a pair of $O(q^2)$ bounds that bracket the value of a selected subset without a solver call. And it contributes a measurement of how much of the observed gain is attributable to selection rather than to the sub-solver behind it, which on the instances studied here turns out to be all of it.

## 3 Methodology

### 3.1 Problem statement

We consider an unconstrained binary quadratic problem over $N$ variables,

$$H(x) = const + h^T x + ½\, x^T J\, x, \quad x \in \{0,1\}^N. \tag{1}$$

with $J$ symmetric and zero on the diagonal. Because $x^2 = x$ for binary variables, a QUBO has no separate square terms: the diagonal of $Q$ is exactly the linear part, which is why $h$ and $J$ suffice.

The operative constraint is that the available sub-solver, a quantum device, or any solver whose cost grows steeply in the number of variables, can only accept $q$ variables at a time, with $q \ll N$. The question is therefore not how to solve a QUBO, but which $q$ of the $N$ variables are worth re-optimizing now.

### 3.2 The sub-QUBO outer loop

The loop is standard and is summarized in Figure 1. Given an incumbent $x$, one round selects a subset $S$ with $|S| \leq q$, clamps the variables outside $S$ at their current values, solves the resulting $q$ -variable sub-problem exactly or heuristically, and writes the solution back. Clamping is not a matter of extracting the $S$-rows and $S$-columns of Q: the fixed variables still act on the free ones through a folded field:

$$d_i = \Sigma \left(Q_{ij} + Q_{ji}\right)\hat{x}j_j, summed\ over\ all\ j \notin S, \tag{2}$$

which shifts the linear coefficients of the sub-problem. Omitting this term silently solves a different problem.

Every step except selection is fixed by the formulation. Selection is where the design freedom lies, and it must be exercised without calling the sub-solver, otherwise the cost of deciding what to solve exceeds the cost of solving it.

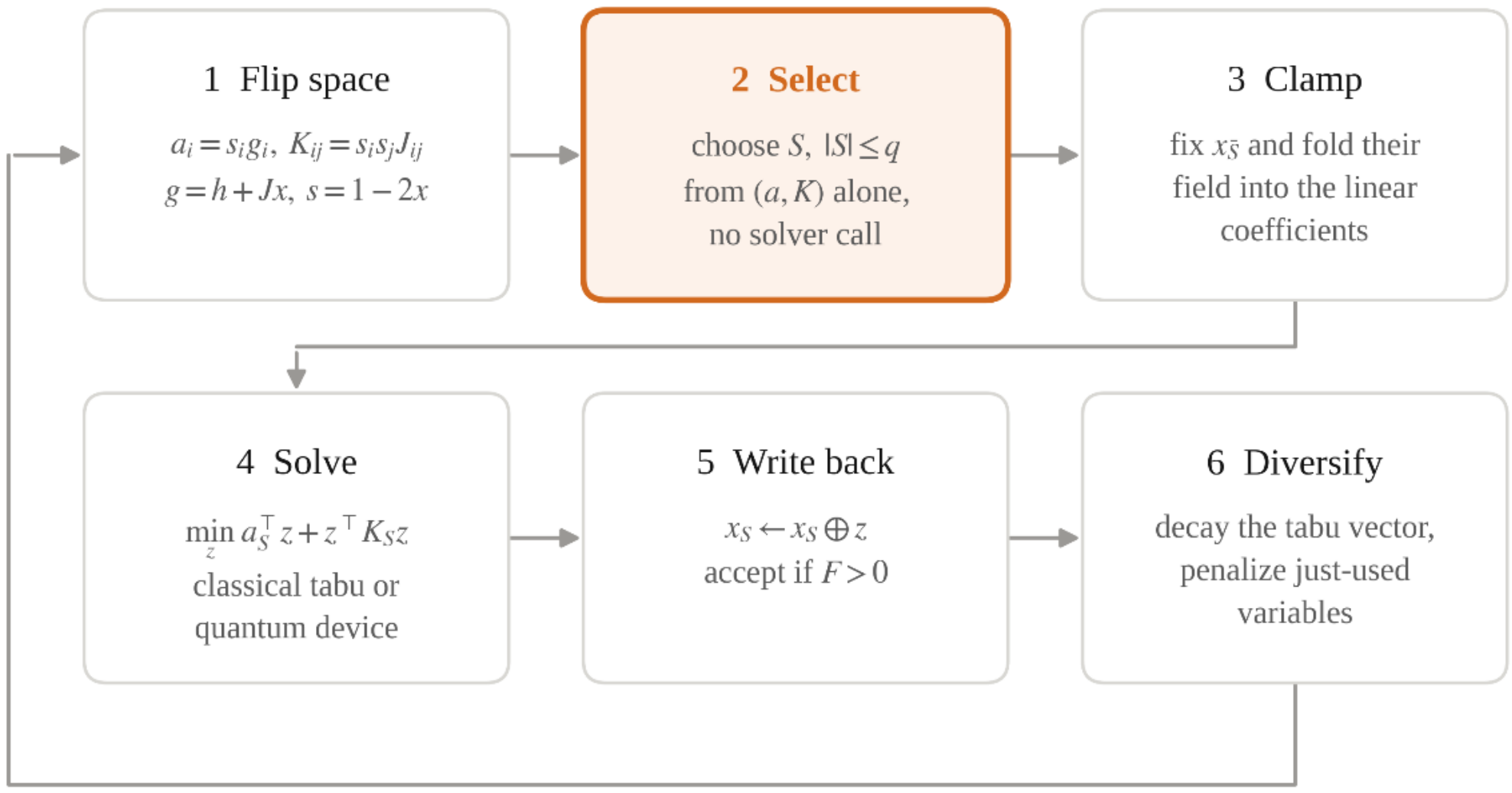

**Figure 1. The sub-QUBO outer loop.** One round of the loop. Steps 1 and 3-6 are determined by the formulation; Step 2: choosing which $q$ of the $N$ variables to release, is the subject of this paper and must be decided from the incumbent and the coupling structure alone, before any sub-problem is solved.

## 3.3 Flip-space reformulation

It is convenient to work in terms of which variables change rather than what they change to. Let $z \in \{0,1\}^N$ indicate the flipped variables, so that the new point is $x' = x + s \cdot z$ with $s = 1 - 2x$. Writing $g = h + Jx$ for the gradient at the incumbent, a direct expansion gives

$$H(x') - H(x) = a^T z + ½\, z^T K\, z, \tag{3}$$

$$a_i = s_i g_i, \tag{4}$$

$$K_{ij} = s_i s_j J_{ij}, \tag{5}$$

This reformulation is what makes selection tractable. It separates the two quantities a rule can read: $a_i$ is the cost (or gain) of flipping variable $i$ on its own, and $K_{ij}$ is the correction that applies when $i$ and $j$ flip together. Both are computed once per round and are available to the selector at no additional solver cost. $K$ inherits the zero diagonal of $J$ .

### 3.4 The computation value of a subset

For a candidate subset $S$ define the computation value in Eq (6):

$$F(S) = -\,min\ over\ z\ with\ supp(z) \subseteq S{:}\ [\, a^T z \ +\ ½\, z^T K\, z\,] \tag{6}$$

The largest objective improvement obtainable by re-optimizing $S$ alone. $F(S) \geq 0$ always, since $z{=}0$ is feasible, and $F$ is monotone: releasing more variables cannot reduce the attainable improvement. The ideal selection problem is max $F(S)$ subject to $|S| \leq q$.

The difficulty is circular. Evaluating $F(S)$ requires solving the very sub-problem whose value we are trying to predict, so a rule that evaluates many candidates exactly defeats its own purpose. Nor is $F$ submodular, so the standard greedy guarantee does not apply: variables can be strictly complementary, contributing nothing individually and a great deal jointly. The example in Figure 2 has $F(\{x_1\}) = F(\{x_2\}) = 0$ but $F(\{x_1, x_2\}) =$ 2.9 increasing, not diminishing, returns. The practical question is therefore whether $F(S)$ can be predicted, or at least usefully bounded, from the QUBO structure alone.

### 3.5 Why impact indexing fails at a local optimum

The most common rule ranks variables by $|\Delta E_i|$, in our notation, by $|a_i|$. The outer loop, however, spends nearly all of its time at points that are already 1-opt optimal, and 1-opt optimality is exactly the statement that $a_i \geq 0$ for every $i$. At such a point, ranking by $|a_i|$ selects the variables whose individual flip is most expensive, while every source of improvement lives in $K$, which the rule never inspects. Figure 2 makes the failure concrete on four variables with $q = 2$. Impact indexing selects $\{x_3, x_4\}$, the two largest $a_i$. Neither flip pays off and the coupling between them is positive, so the sub-solver correctly returns $z = 0$ and the round yields nothing. A rule that reads $K$ instead selects $\{x_1, x_2\}$, where $K_{12} = -4$ outweighs the two small individual costs, and obtains an improvement of 2.9 from the same budget and the same solver. A second, subtler consequence is that the $|a_i|$ ranking is deterministic. Consecutive rounds therefore propose almost identical subsets, and an external diversification mechanism becomes necessary rather than optional, a prediction we test directly in Section 5.5.

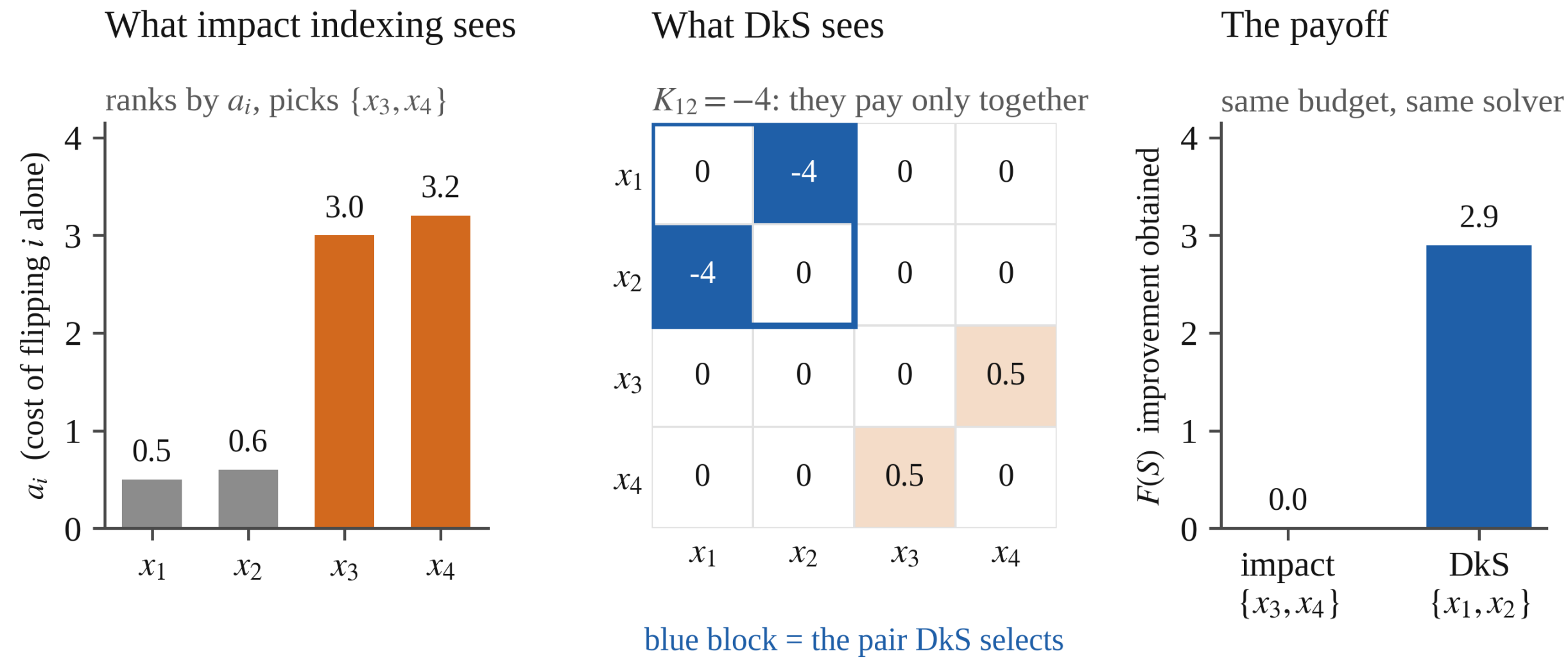

**Figure 2. What each selection rule can see.** A four-variable instance at a 1-opt optimum with $q = 2$. Left: the linear costs $a_i$, which are all that impact indexing reads. Middle: the coupling matrix $K$; the negative entry $K_{12} = -4$ means $x_1$ and $x_2$ pay off only together. Right: the improvement actually obtained.

### 3.6 The DkS selector

Consider the specific candidate solution that flips every variable in a subset $T$. Its improvement is

$$-\left[\sum_{i\in T} a_i + \sum_{i<j\in T} K_{ij}\right], \tag{7}$$

which is a lower bound on $F(T)$. Maximizing this bound, and relaxing $K_{ij}$ to its negative part $\left[K_{ij}\right]^{-} = \max\left(0, -K_{ij}\right)$ so that the edge weights are non-negative, yields

$$\max \mathit{over}\ |T| \leq q\ \mathit{of}\ \left[\sum_{i<j\in T} \left[K_{ij}\right]^{-} - \sum_{i\in T} a_i\right], \tag{8}$$

This is a prize-collecting densest-k-subgraph problem: find $q$ nodes that are densely connected by strong negative couplings while paying as little as possible in individual flip costs. We solve it greedily. The seed is the best pair, after which variables are added one at a time by the incremental score $\sum_{i<j\in T} \left[K_{ij}\right]^{-} - \sum_{i\in T} a_i$ . The cost is $O(qN)$ er round, negligible beside a single sub-solver call.

One detail in the seeding step is easy to get wrong and materially affects small instances. The value of a candidate pair is not only the "flip both" option, because "flip just one" is also available and is sometimes the better move. The seed score must therefore be

$$M_{ij} = \mathit{max}\left(-a_i - a_j - K_{ij}, -a_i, - a_j\right), \tag{9}$$

and taking only the first term causes the greedy to stall at inferior subsets whenever no pair is jointly profitable.

### 3.7 A selection-quality certificate

Because the selector never calls the sub-solver, it is useful to bracket what it has selected. Two bounds are available in $O(q^2)$:

$$L(S) = \mathit{max}\left(0, \mathit{max\ over\ i\ of}\,(-a_i), \mathit{max\ over\ i<j\ of}\ (-a_i, - a_j - K_{ij},)\right), \tag{10}$$

$$U(S) = \sum_i [a_i]^- + \sum_{i<j} \left[K_{ij}\right]^-. \tag{11}$$

so that $L(S) \leq F(S) \leq U(S)$. The lower bound holds because each of its arguments corresponds to a feasible $z$. The upper bound holds because dropping every non-negative contribution from the objective can only increase the attainable improvement. Neither requires a solver call, whereas $F(S)$does. Section 5.2 reports how tight the bracket is in practice, and shows that $U$ alone is a poor guide.

### 3.8 Diversification

A tabu vector $t$, decayed by a factor 0.8 each round and incremented on the variables just used, is added to a as a penalty $\varepsilon t$ before selection. This is the only mechanism by which one round influences the next. Whether it is essential or merely convenient turns out to depend entirely on the selection rule, and is measured in Section 5.5.

Written out, the mechanism is as follows. Let $\tau(t)$ denote the tabu vector at the start of round $t$, initialized to zero, and let $1S(t)$ be the indicator vector of the subset selected in that round. The vector is then updated as:

$$\tau^{(t+1)} = \rho\,\tau^{(t)} + \mathbf{1}_{S_t} \tag{12}$$

with decay factor $\rho = 0.8$. Selection in round $t$ does not read $a$ directly, but the penalized coefficients of Eq (13):

$$\tilde{a}^{(t)} = a^{(t)} + \varepsilon\,\tau^{(t)} \tag{13}$$

where $\varepsilon > 0$ fixes the penalty strength and is held constant across all experiments. A variable selected in the immediately preceding round therefore carries a penalty of $\varepsilon$, one selected $r$ rounds earlier carries $\varepsilon\rho r$, and a variable never selected carries none. Because $\rho < 1$ the penalty decays geometrically, and the loop retains a memory of roughly $1/(1-\rho) = 5$ rounds.

Both selection rules consume $\tilde{a}$ in place of $a$, so the mechanism is orthogonal to the choice of rule, which is what allows Section 5.5 to vary the two independently. The time constant also gives a reading of the stopping rule recommended in Section 6: a run of fewer than about five unproductive rounds is not long enough for the penalty to have redirected selection onto variables not yet examined, which is consistent with the recovery at round 15 reported in Section 5.6.

### 3.9 The quantum sub-solver interface

The loop reaches the sub-solver through a single contract: given $(S, a_S, K_S)$, return the improvement $F$ and the flip vector z . Nothing else in the method changes when the backend changes, which is what makes the null ablation of Section 5.6 possible, the identical selection trace can be replayed against a classical and a quantum sub-solver.

The quantum backend is Kipu Quantum's Iskay optimizer (bias-field digitized counter-diabatic quantum optimization) running on ibm_rensselaer, whose configuration is given in Section 4.4, with a 1:1 mapping from variables to programmable qubits. The Iskay service itself accepts up to 156 variables, so the binding constraint here is the device, not the interface. Because hardware calls are slow and metered, every call is keyed by a hash of $(S, a_S, K_S)$ and logged write-ahead to disk, so an interrupted run resumes without repeating work and repeated sub-problems are not re-submitted.

## 4 Experimental setup

### 4.1 Instances

Two real city networks from the Transportation Networks for Research repository, in TNTP format: Chicago-Sketch with $n = 387$ traffic analysis zones, and Philadelphia with $n = 1{,}525$. Both have measured OD matrices with heavy-tailed coupling strengths,  the coefficient of variation of $|K_{ij}|$ is 2.92 and 6.05 respectively.

For contrast we also report on the synthetic instance the study began with, whose coupling matrix is near-dense and low-rank: the balance term carries 99.9% of the total coupling mass and the CV of $|K|$ is only 0.58. On such an instance no structure-reading rule can outperform random selection, and none does. We report it rather than omit it, and Section 6 turns it into a diagnostic to run before applying the method.

### 4.2 Objective

Zone bipartition: assign $n$ zones to two regions so as to minimize the OD flow cut between them, subject to a spatial-coherence penalty of weight $\gamma$ and a size-balance constraint. The formulation follows (Ke et al., 2026), so that the comparison isolates the selection rule rather than confounding it with a change of objective. Unless stated otherwise $\gamma = 1.0$. Adjacency is defined either by Delaunay triangulation of zone centroids (the geometric proxy) or by road-network first-ring adjacency, described in Section 5.4.

### 4.3 Protocol

Every comparison is at equal cost. Within any one experiment the call budget and the number of random starts are the same for every selection rule, and the value reported is the mean over starts. The budget itself differs between experiments: the γ and q sweeps of Section 5.1 use 60 rounds, while the cost comparison of Section 5.3 and the ablations of Section 5.4 and Section 5.5 use 40. The sub-solver is exhaustive enumeration for $q \leq 16$ and multi-start tabu search in flip space for $q > 16$, and the same sub-solver is applied to every rule within an experiment.

Different experiments vary different factors, so the objective values in one table are not in general comparable with those in another. Every table below therefore carries its configuration in full, instance, adjacency definition, $\gamma$, sub-problem size, call budget and number of starts, and absolute values should be compared only between rows that share a configuration. The comparisons the paper draws conclusions from are always within a single table, between rules that differ in exactly one respect.

### 4.4 Quantum hardware and service

The quantum sub-solver runs through the Iskay Quantum Optimizer, a Qiskit Function provided by Kipu Quantum Kipu Quantum (2026), on the IBM device ibm_rensselaer. Table 1 lists the device characteristics reported by the platform at the time of the runs. Two of them matter for what follows.

**Table 1.** Characteristics of ibm_rensselaer as reported on the IBM Quantum platform at the time of the experiments. Devices are recalibrated regularly, so the error rates are a snapshot rather than a fixed specification.

| Property | Value | Property | Value |
|---|---|---|---|
| Processor type | Nighthawk r1 | Median 2Q (CZ) error | $5.20 \times 10^{-3}$ |
| Programmable qubits | 120 | Best 2Q error | $1.15 \times 10^{-3}$ |
| Physical qubits | 338 | Layered 2Q error | $1.44 \times 10^{-2}$ |
| Basis gates | cz, id, rx, rz, sx, x | Median SX error | $2.76 \times 10^{-4}$ |
| CLOPS | 270 000 | Median readout error | $8.06 \times 10^{-3}$ |
| Max circuits per second | 3.8 kHz | Median initialization error | $4.16 \times 10^{-3}$ |
| QPU version | 2.0.1 | Median $T_1$ | 281 µs |
| Region | Washington DC (us-east) | Median $T_2$ | 132 µs |

The first is the distinction between programmable and physical qubits. The device carries 338 physical qubits but exposes 120 programmable ones, and it is the latter that bounds the sub-problem size. Every reference to a 120-qubit device in this paper, including the sub-problem size $q = 120$ at which compilation fails, refers to programmable qubits under the 1:1 variable-to-qubit mapping of Section 3.9.

The second is the gap between the median two-qubit error of $5.20 \times 10^{-3}$ and the layered two-qubit error of $1.44 \times 10^{-2}$, which is almost three times larger. The layered figure is measured with gates executed in parallel, so it reflects the crosstalk and scheduling pressure that arise when many two-qubit gates must run simultaneously. A sub-problem with more coupling terms requires more two-qubit gates, and those gates must be routed and parallelized by the compiler. This is the same pressure that Section 5.7 observes from the outside, where compilation time grows almost as the square of the term count while the number of qubits stays fixed.

# 5 Results

## 5.1 Selection dominates

Figure 3 sweeps the sub-problem size $q$ from 16 to 64 on Philadelphia. The ordering of the four rules is preserved throughout, but the informative comparison is vertical rather than horizontal.

**Table 2.** Final objective on Philadelphia, $n$ = 1,525. Configuration: Delaunay adjacency, $\gamma = 1.0$, 60 rounds, mean ± one standard deviation over 8 random starts. Lower is better.

| $q$ | DkS | random | impact-rolling | Impact\| $\|dE\|$ |
|---|---|---|---|---|
| 16 | 0.068 ± 0.009 | 0.332 ± 0.009 | 0.498 ± 0.016 | 0.456 ± 0.011 |
| 32 | 0.067 ± 0.007 | 0.238 ± 0.011 | 0.250 ± 0.009 | 0.431 ± 0.009 |
| 64 | 0.070 ± 0.009 | 0.159 ± 0.011 | 0.259 ± 0.008 | 0.392 ± 0.009 |

DkS at $q = 16$ reaches 0.068 ± 0.009, while random at $q = 64$ reaches 0.159 ± 0.011. A fourfold increase in device capacity does not buy back a better selection rule. On near-term hardware the return on a better selection rule is far larger than the return on a few dozen more qubits.

The DkS curve is flat rather than monotone. Its three means span 0.003, which is less than half the standard deviation at any single size, so the sweep does not resolve an ordering among them and none should be read from the figure. Two controls support that reading. Every point averages 8 random starts rather than 3, and the bands in Figure 3 are ±1 standard deviation. Rerunning $q = 32$ and $q = 64$ with a much stronger sub-solver, a 24-restart tabu search of 2,500 iterations per restart, moves the mean at $q = 32$ only from 0.067 to 0.066, so the flatness is not an artifact of larger sub-problems being solved less exactly. Under a fixed call budget a larger $q$ means each call covers more variables while the number of rounds is unchanged, and the flat curve says that at $q = 16$ the rule already captures the improvement reachable within that budget.

Reading Table 2 across columns rather than down them raises a question of protocol. As stated in Section 4.3, the sub-solver is exhaustive enumeration for $q \leq 16$ and multi-start tabu search above it, so the $q = 16$ column is the only one in which sub-problems are solved to optimality, and a comparison across $q$ would otherwise vary two things at once. We therefore ran the direct control: DkS at $q = 16$ driven by the same tabu search used at larger $q$, with everything else held identical. The two sub-solvers agree exactly, returning the same objective on every random start, because at $q = 16$ the tabu search already reaches the optimum of every sub-problem it is handed. The choice of sub-solver therefore contributes nothing to the comparison, and the margin reported in Table 2 is attributable to selection alone.

The same observation explains a result reported later. In Section 5.6 the quantum backend returns sub-problem optima identical to exhaustive enumeration on all 42 hardware calls, which is likewise a statement about how easy $q = 16$ sub-problems are rather than about the backend.

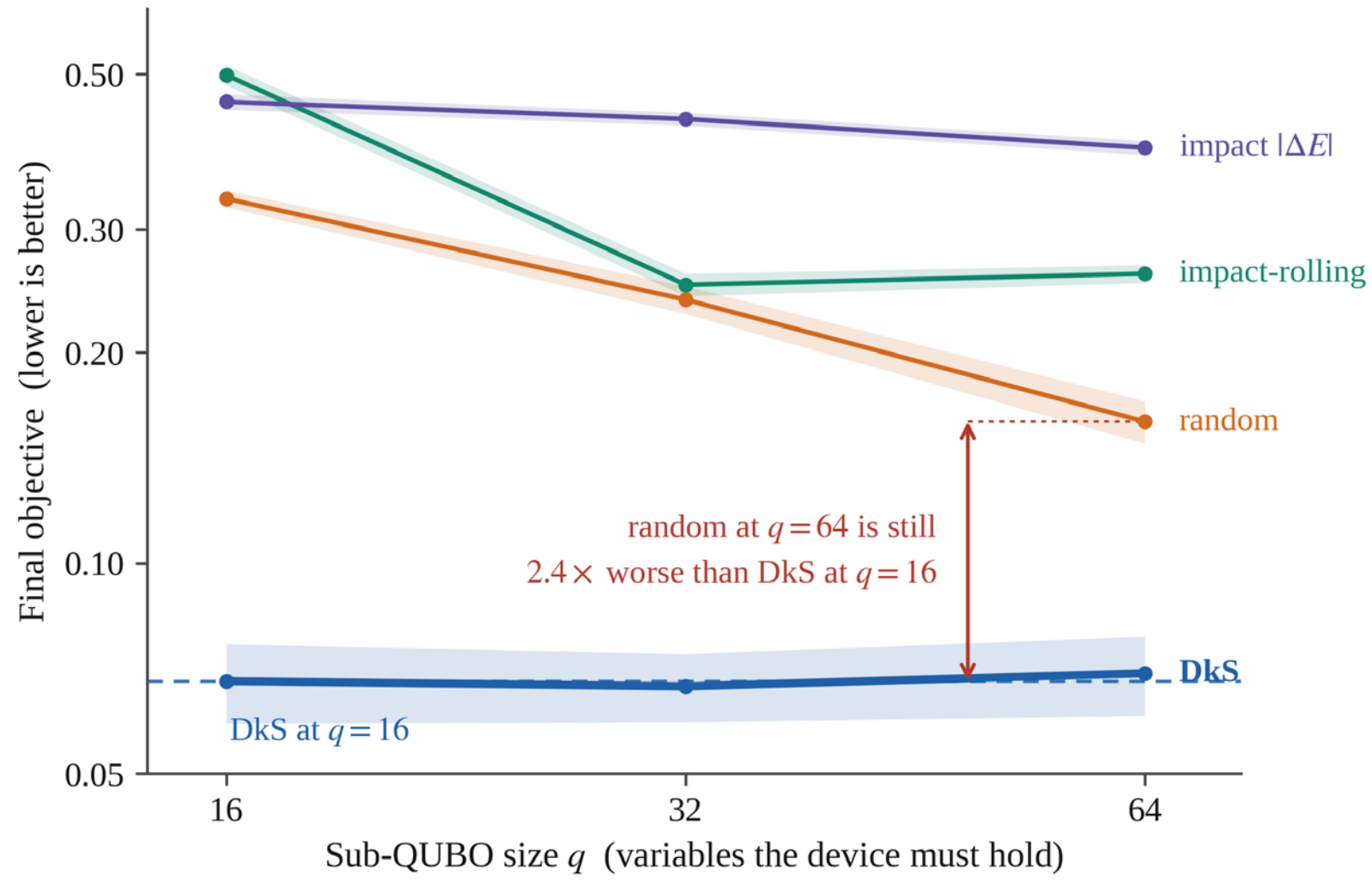

**Figure 3. Better selection substitutes for qubits.** Final objective against sub-problem size on Philadelphia, at a fixed budget of 60 sub-QUBO calls. Lines are means over 8 random starts and bands are ±1 standard deviation. The dashed reference line is DkS at $q = 16$. Random selection at $q = 64$ stays well above that line, and the DkS band overlaps it at every size, so the rule is flat in $q$ rather than best at $q = 16$.

The conclusion is robust in $\gamma$. Figure 4 sweeps $\gamma$ from 0.1 to 3.0 on both instances, moving the adjacency term's share of total coupling mass from 7% to 68% on Chicago and 6% to 67% on Philadelphia. The four curves are approximately parallel on log-log axes and the ordering never changes.

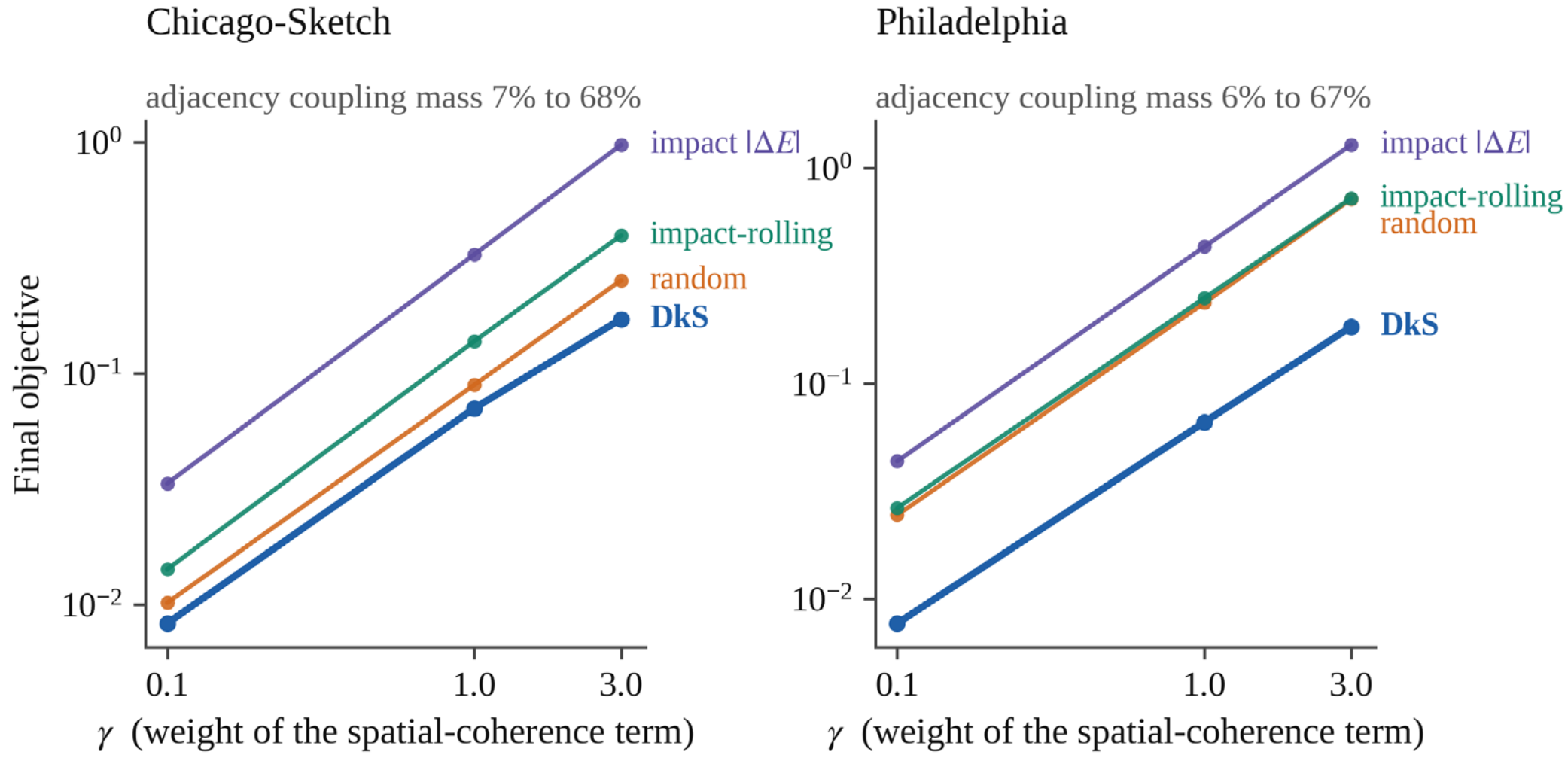

**Figure 4. The ranking is stable across the spatial-coherence weight.** Final objective against $\gamma$ on both instances, at Delaunay adjacency, $q = 32$, 60 rounds and 5 random starts.

## 5.2 Convergence and the certificate

In Figure 5 (left) DkS reaches its final level within roughly five rounds, while random is still improving at round 40 and impact $|dE|$stalls completely at round 3, the deterministic-ranking failure predicted in Section 3.5.

Figure 5 (right) plots the certificate of Section 3.7. During the first five rounds $F$ sits inside a narrow band between $L$ and $U$. After the incumbent converges, $F$ drops to zero while $U$ continues to spike repeatedly. The upper bound alone therefore badly overestimates the remaining improvement, and a practical stopping rule should be built on $L$, not $U$.

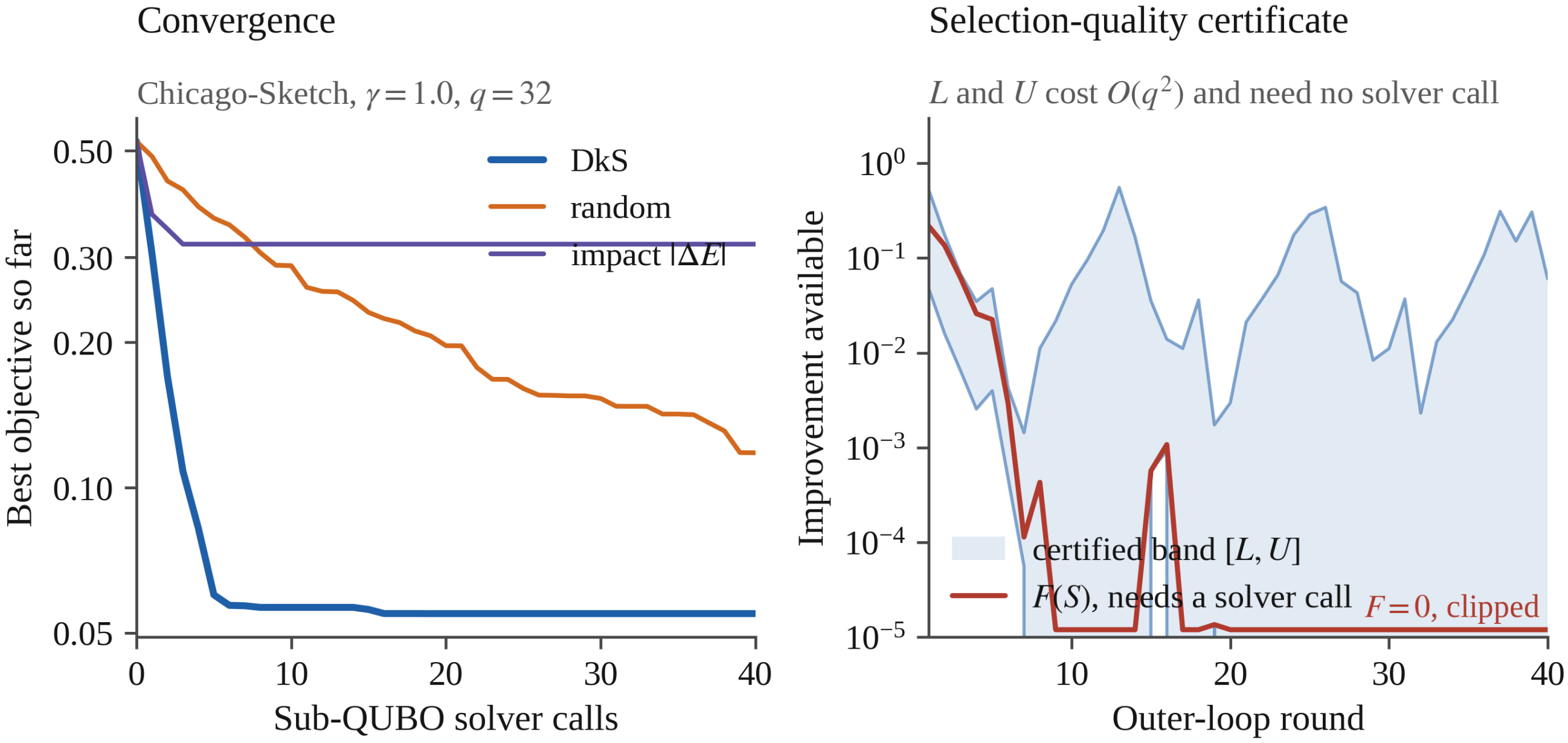

**Figure 5. Convergence and selection-quality certificate.** Left: best objective so far against sub-QUBO solver calls, Chicago-Sketch, $\gamma$ = 1.0, $q$ = 32. Right: the certified band $L(S) \leqslant F(S) \leqslant U(S)$ for the DkS selector. $L$ and $U$ cost O(q²) and need no solver call; $F$ requires one.

The difference is visible on the ground. Figure 6 draws the bipartition of both instances in geographic coordinates, with every zone rendered as the Voronoi cell of its centroid so that the 1,525-zone instance is as legible as the 387-zone one. On Chicago-Sketch the impact-rolling solution is fragmented across the map while the DkS solution resolves into two connected regions, and the cut origin-destination flow falls from 316,326 to 243,132, or from 25% to 19% of all demand. On Philadelphia the same contrast holds, with cut flow falling from 5,819,172 to 4,274,013, or from 31% to 23%.

Spatial coherence is reported as a number rather than left to the eye. Moran's $I$ of the region label over the Delaunay adjacency graph rises from +0.71 to +0.76 on Chicago-Sketch and from +0.60 to +0.75 on Philadelphia, while the share of adjacency edges cut falls from 13.3% to 11.4% and from 20.2% to 12.6%. Under DkS the two instances reach the same level of coherence despite a fourfold difference in size, which is the point that matters: coherence follows from the selection rule and is not a property of the smaller instance. For transport planning only the DkS partitions are deliverable.

**(a) Chicago-Sketch,** $n = 387$

impact-rolling

cut OD flow 316,326 (25%)

objective 0.1156 Moran's $I$ +0.71

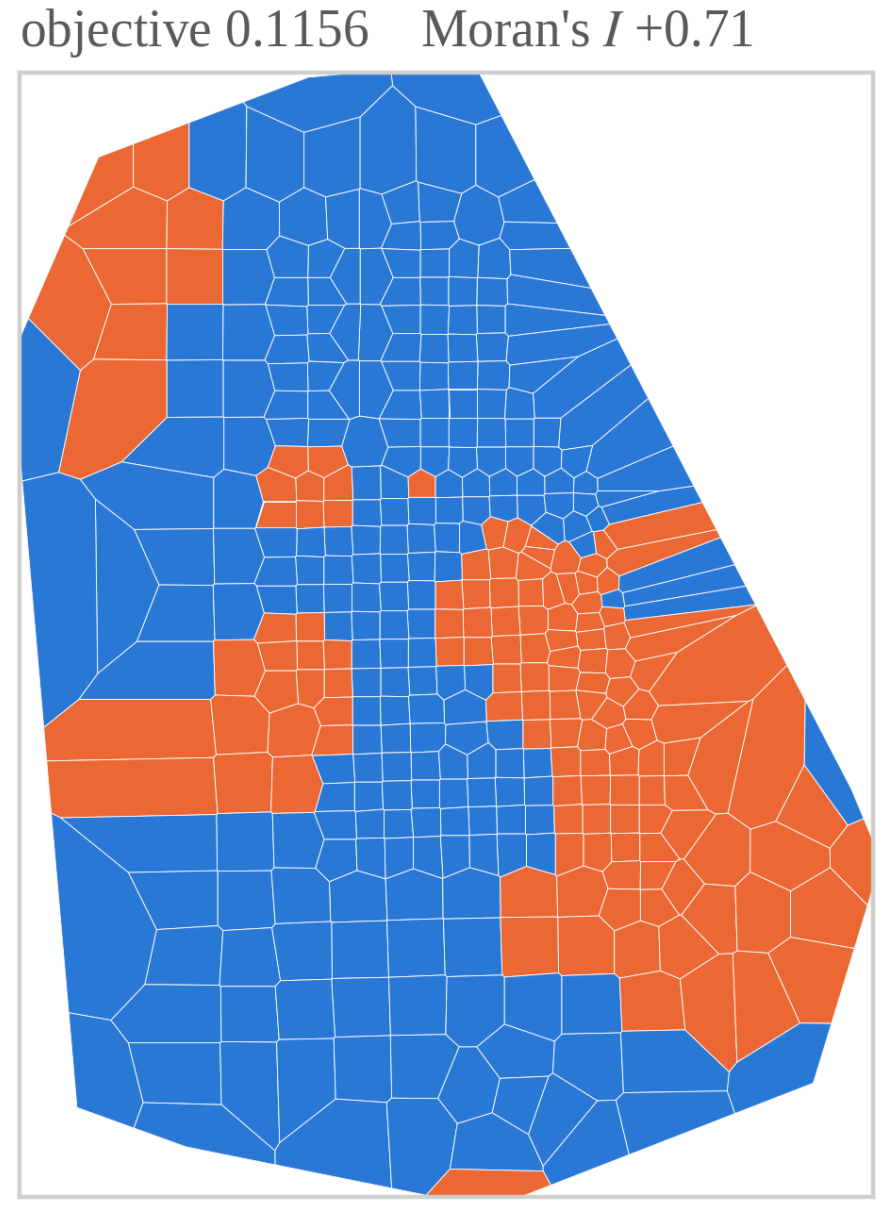

DkS

cut OD flow 243,132 (19%)

objective 0.0454 Moran's $I$ +0.76

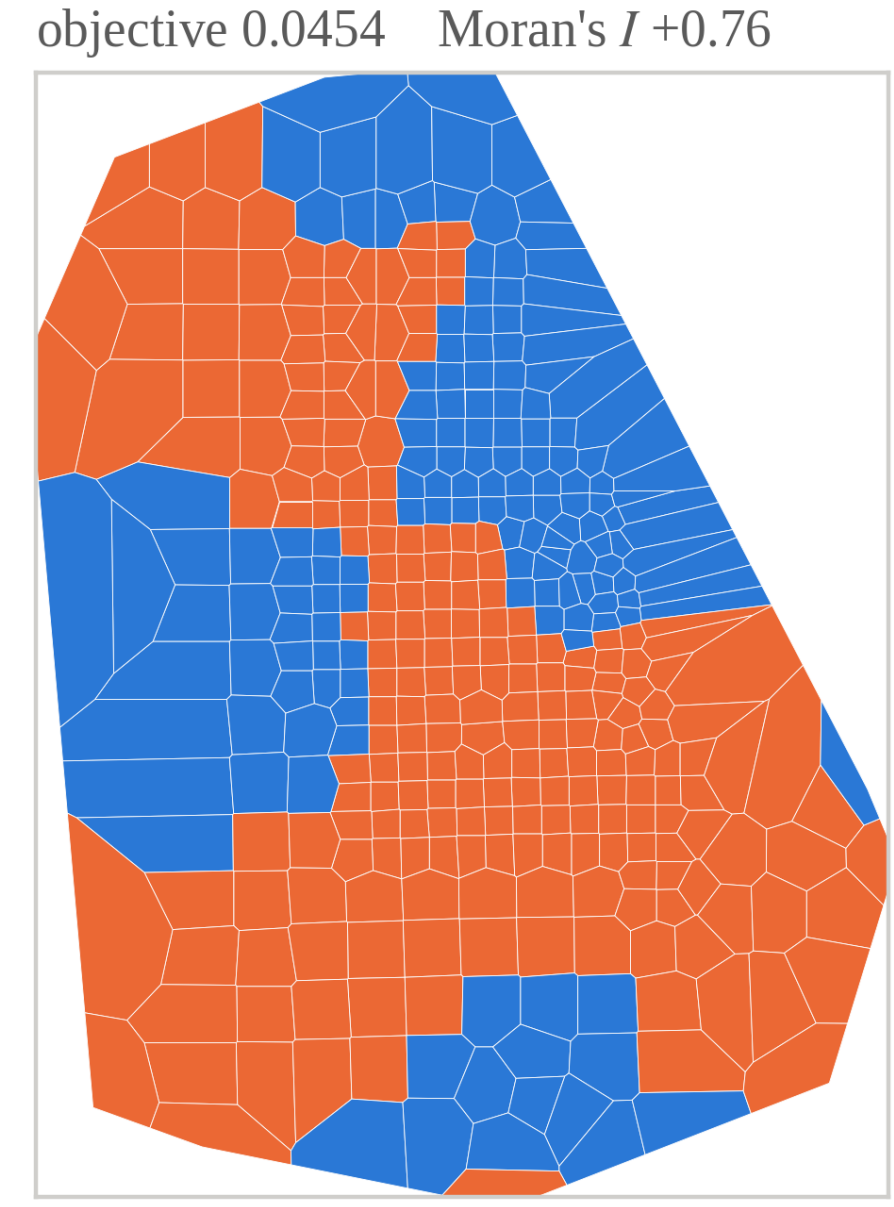

**(b) Philadelphia,** $n = 1525$

impact-rolling

cut OD flow 5,819,172 (31%)

objective 0.1857 Moran's $I$ +0.60

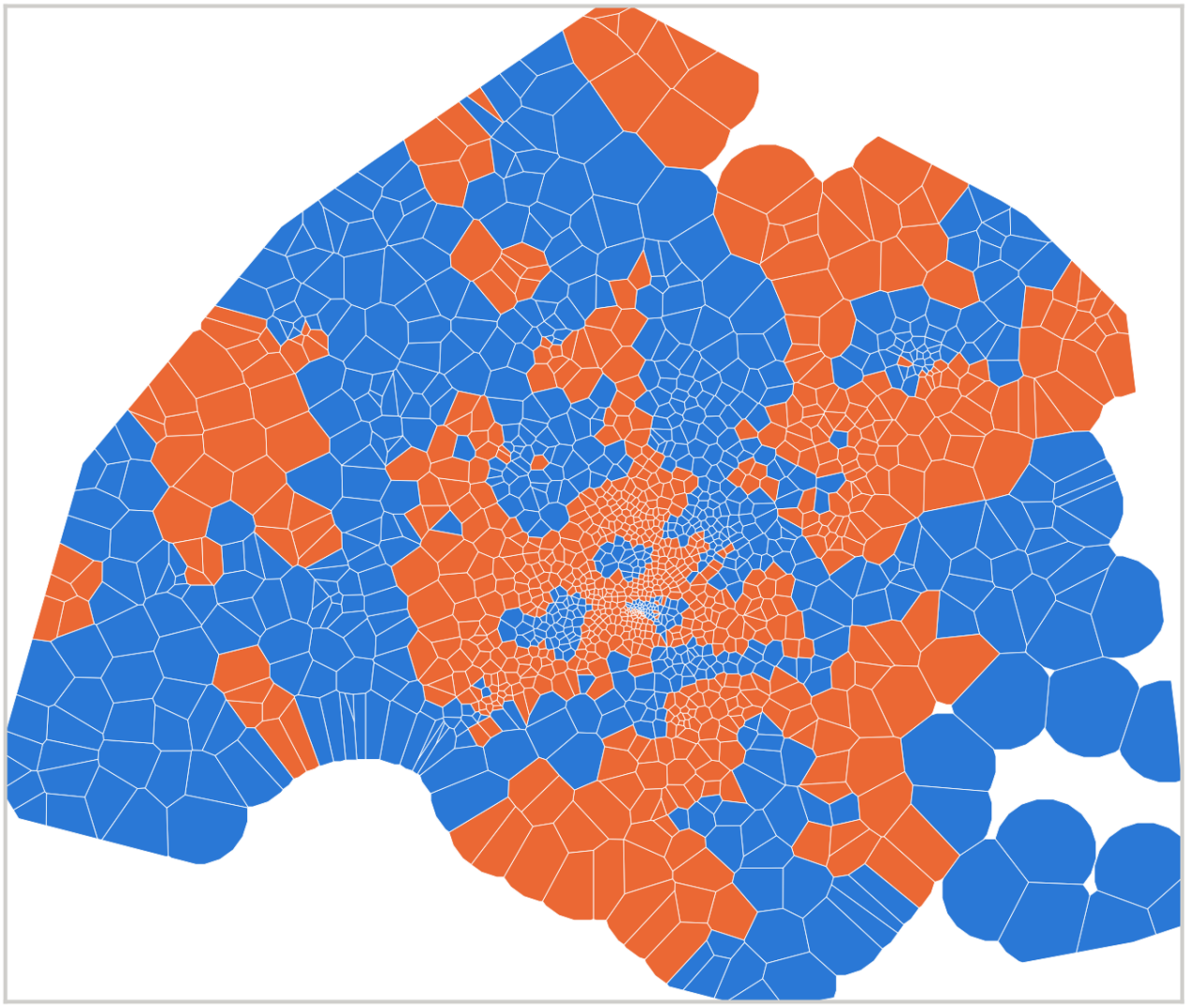

DkS

cut OD flow 4,274,013 (23%)

objective 0.0560 Moran's $I$ +0.75

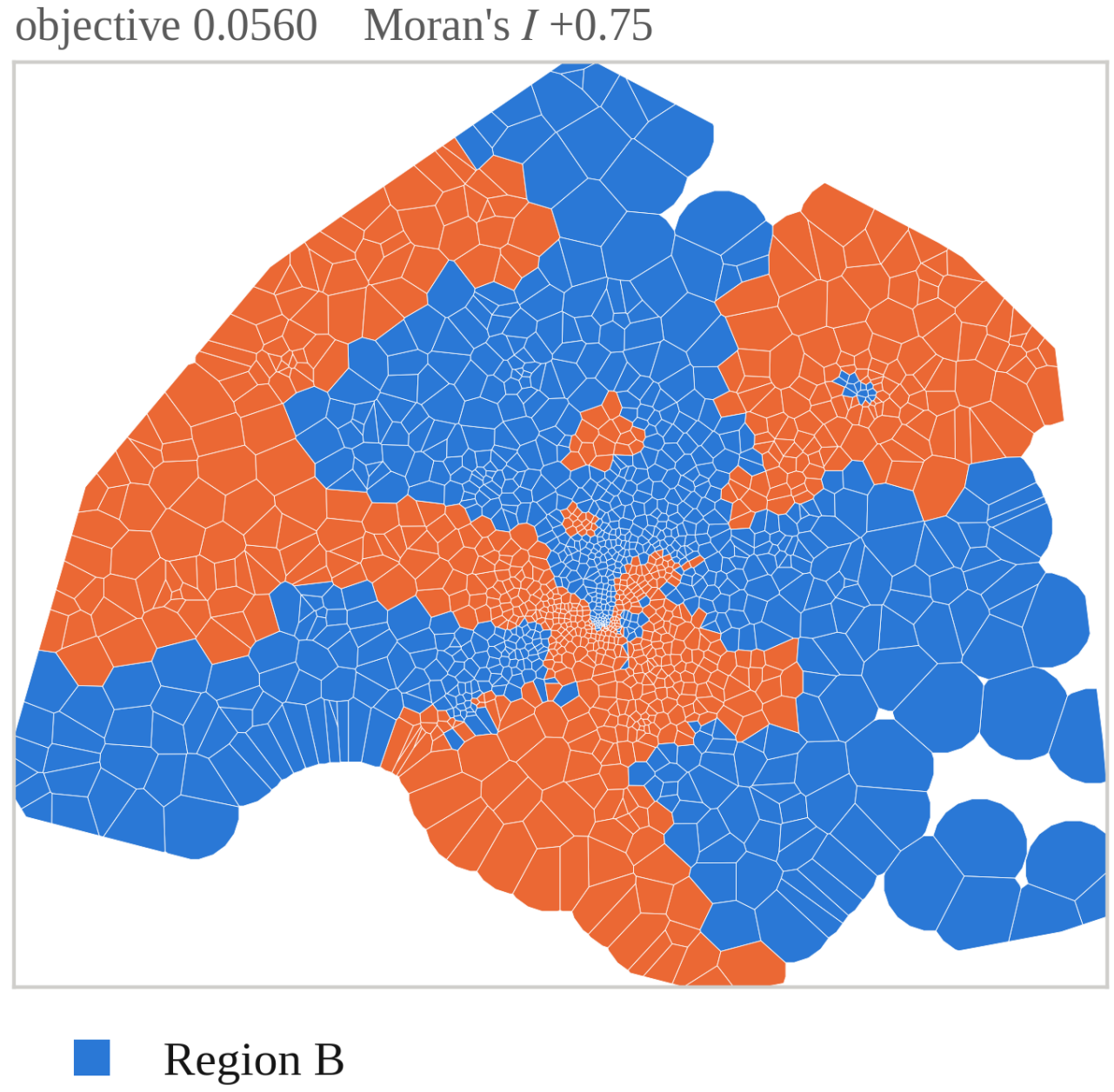

Region A Region B

**Figure 6. The partitions differ in kind, not only in objective value.** Zone bipartitions under two selection rules, at Delaunay adjacency, $\gamma = 1.0$, $q = 32$, 200 rounds. (a) Chicago-Sketch, $n = 387$. (b) Philadelphia, $n = 1{,}525$. Cut OD flow is the total origin-destination demand whose two endpoints fall in different regions. Moran's $I$ is computed on the Delaunay adjacency graph, where +1 is perfect spatial clustering and 0 is a random labelling.

### 5.3 Equal-cost comparison against sample-and-solve

A reviewer will reasonably ask why one should not simply sample 20 candidate subsets per round, solve each, and keep the best. That baseline spends 20× the sub-solver calls. Table 3 reports quality against cost for every method, and Figure 7 plots the same data with cost on the horizontal axis and quality on the vertical axis, so that bottom-left is better.

**Table 3.** Quality against cost. Configuration: Delaunay adjacency, $\gamma = 1.0$, $q = 32$, 40 rounds, 3 random starts on Chicago-Sketch and 2 on Philadelphia. DkS matches or beats sample-and-solve at one twentieth of the cost. At matched cost, sample-and-solve is not competitive.

| Method | Calls | Chicago | Philadelphia |
|---|---|---|---|
| DkS | 40 | 0.078 | 0.055 |
| L-shortlist (M = 20) | 40 | 0.062 | 0.199 |
| sample&solve 20× | 800 | 0.081 | 0.177 |
| sample&solve, cost-matched | 40 | 0.39 | 0.42 |

$L$ -shortlist is a variant that uses the lower bound $L(S)$ to shortlist M candidates before solving. It improves Chicago further but degrades Philadelphia badly, and we do not propose it as a recommended method. We tested the natural explanation, that $L$ stops ranking candidates faithfully on the larger instance, and refuted it: the Spearman correlation between $L$ and the true $F$ is 0.836 on Chicago and 0.784 on Philadelphia, and the regret from selecting by $L$ is in fact lower on Philadelphia (5.9% against 14.0%). The reversal therefore has some other cause, which we report as an open observation rather than explain.

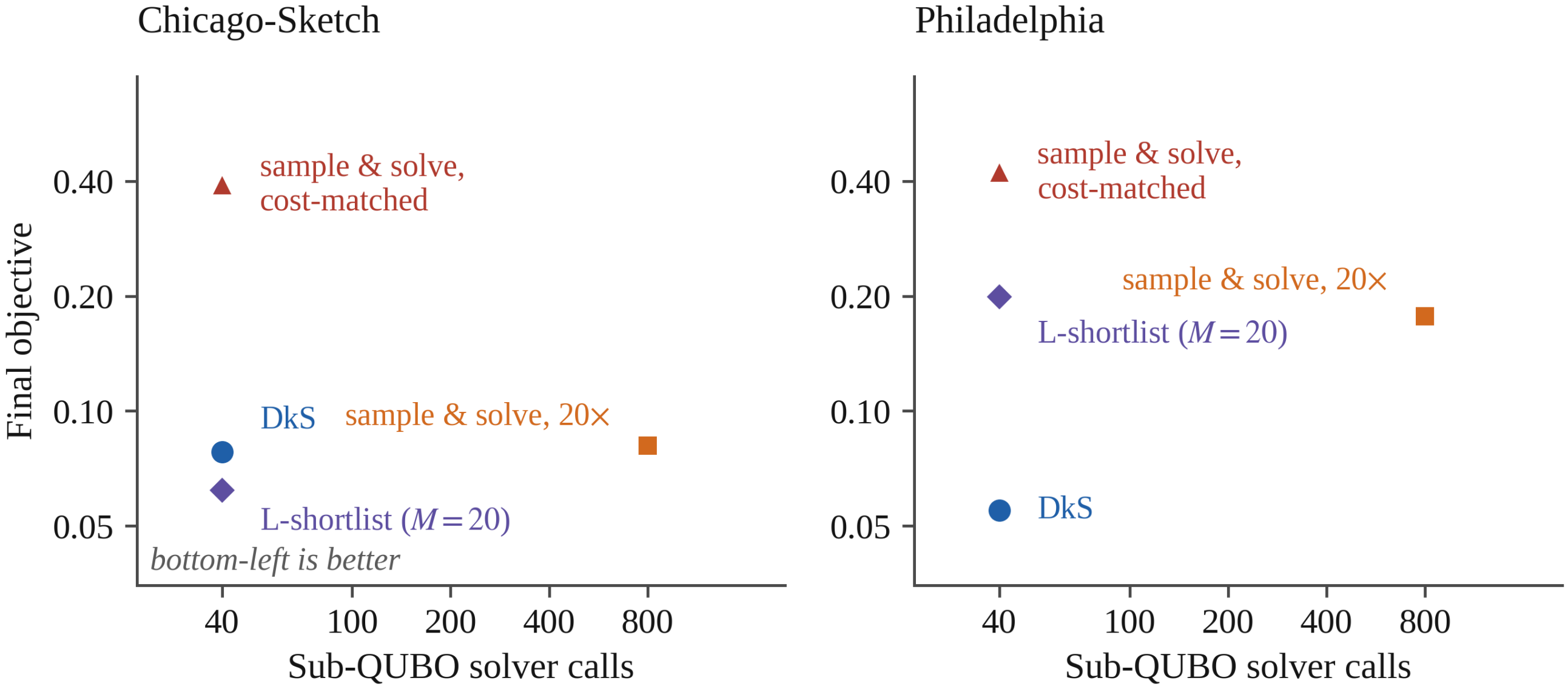

**Figure 7. Efficiency frontier against sample-and-solve.** Objective against number of sub-QUBO solver calls on both instances, where bottom-left is better. DkS is the rule proposed here. L-shortlist is a variant examined in the text but not recommended, for the reason given there.

## 5.4 Transport structure entering the selection

Delaunay triangulation of zone centroids is a geometric proxy: rivers, limited-access freeways and rail corridors all sever pairs of zones that are geometrically adjacent. We replace it with road-network first-ring adjacency, obtained by breadth-first search outward from each zone centroid on the actual road network, stopping at any other centroid. Table 4 reports how far the two definitions diverge.

**Table 4.** Edge counts under the two adjacency definitions. Overlap is the fraction of road first-ring edges that also appear in the Delaunay triangulation.

| Instance | Delaunay edges | Road first-ring edges | Overlap |
|---|---|---|---|
| Chicago-Sketch | 1148 | 689 | 58.3% |
| Philadelphia | 4559 | 1880 | 38.2% |

The two adjacency definitions produce different QUBO instances, so absolute objective values are not comparable across them. The meaningful quantity is the DkS-to-random ratio within each adjacency, which Table 5 reports.

**Table 5.** Configuration: $\gamma = 1.0$, $q = 32$, 40 rounds, 3 random starts, identical in both columns so that only the adjacency definition changes. Switching to road-network adjacency widens the relative advantage of DkS by a factor of roughly 1.6 to 1.9 on both instances.

| Instance | Delaunay: DkS / random | Road first-ring: DkS / random |
|---|---|---|
| Chicago-Sketch | 0.0753 / 0.1068 = 0.705 | 0.0431 / 0.0989 = 0.436 |
| Philadelphia | 0.0699 / 0.2846 = 0.246 | 0.0296 / 0.2258 = 0.131 |

The proposed explanation is that road adjacency is considerably sparser than Delaunay, which concentrates the coupling structure and so makes it more discriminative, and that DkS benefits because it reads that structure while random selection cannot. Table 6 tests the explanation directly. It reports the coefficient of variation of $|\boldsymbol{K}|$ introduced in Section 4.1, together with two statistics computed on the negative part of $K$, which is the part the selector actually reads: its coefficient of variation, and the share of the total negative coupling mass carried by the strongest one per cent of entries.

**Table 6.** Concentration of the coupling structure under the two adjacency definitions, measured at a 1-opt optimum with $\gamma = 1.0$. The last column is the fraction of total negative coupling mass carried by the strongest one per cent of entries. Higher values mean a more discriminative structure.

| Instance and adjacency | Adj. edges | CV of $\|\boldsymbol{K}\|$ | CV of $\|\boldsymbol{K}\|^{-}$ | Top 1% share |
|---|---|---|---|---|
| Chicago-Sketch, Delaunay | 1148 | 6.03 | 5.16 | 41.1% |
| Chicago-Sketch, road first-ring | 689 | 7.66 | 6.64 | 50.3% |
| Philadelphia, Delaunay | 4559 | 11.34 | 10.82 | 59.4% |
| Philadelphia, road first-ring | 1880 | 17.13 | 17.66 | 64.2% |

Every measure moves in the predicted direction. On Chicago-Sketch the coefficient of variation of the negative couplings rises from 5.16 to 6.64, and the share of negative coupling mass carried by the strongest one per cent of entries rises from 41.1 to 50.3 per cent. On Philadelphia the same two quantities rise from 10.82 to 17.66 and from 59.4 to 64.2 per cent. The effect is ordered consistently across the two instances as well: Philadelphia shows the larger increase in concentration, 63 per cent against 29 per cent, and it is also the instance on which the advantage of DkS widens the most, by a factor of 1.88 against 1.62. The mechanism offered for the result is therefore measured rather than assumed.

One distinction this measurement does not draw. It establishes that the sparser adjacency concentrates the coupling structure, which is the claim made here, but it cannot separate that claim from the stronger one that road-network topology carries information which an equivalent amount of arbitrary sparsification would not. Deleting a matching number of Delaunay edges at random and repeating the measurement would settle the question, and we have not done so.

Figure 8 shows the same comparison of objective values graphically. In Table 6, this is a concrete instance of domain knowledge translating into algorithmic gain, rather than a general-purpose solver applied to transport data.

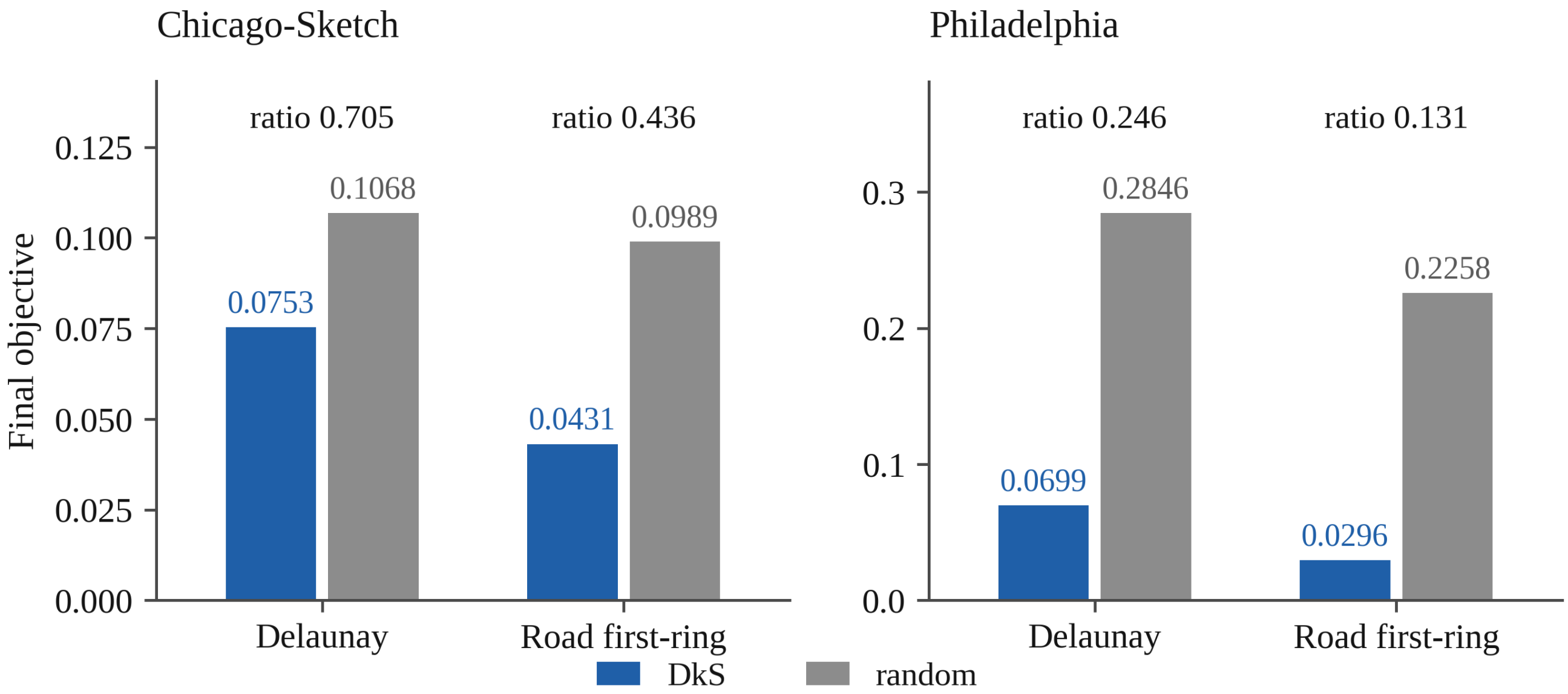

*Absolute levels are not comparable across the two adjacency definitions, which are different QUBO instances. The ratios are.*

**Figure 8. Real road connectivity instead of centroid Delaunay.** Final objective under both adjacency definitions. Absolute values are not comparable across the two definitions because they are different QUBO instances; the within-definition ratios in Table 5 are.

## 5.5 Selection rule versus diversification mechanism

In the results above DkS used a tabu penalty while impact used a rolling window, confounding two factors. Table 7 and Figure 9 report the full $2 \times 3$ ablation that separates them.

**Table 7.** Configuration: road first-ring adjacency, $\gamma = 1.0$, $q = 32$, 40 rounds, 3 random starts, the same run that produced Table 5, so the tabu column reproduces the road-adjacency column of Table 5 exactly. Random-selection baselines are 0.0989 (Chicago) and 0.2258 (Philadelphia).

| Rule | none | rolling | tabu | worst / best |
|---|---|---|---|---|
| Chicago — impact | 0.3097 | 0.0745 | 0.0794 | 4.16× |
| Chicago — DkS | 0.0429 | 0.0402 | 0.0431 | 1.07× |
| Philadelphia — impact | 0.3996 | 0.0641 | 0.0830 | 6.23× |
| Philadelphia — DkS | 0.0312 | 0.0383 | 0.0296 | 1.29× |

Almost all of impact's performance comes from the bolt-on diversification mechanism: without it, impact is three to four times worse than random selection. DkS is nearly insensitive to the mechanism, the spread across the three settings is 7% to 29%, and even with no diversification at all it outperforms the best impact configuration.

This is the prediction of Section 3.5 confirmed. The DkS objective contains $\sum K_{ij}^{-}$ explicitly and therefore favors mutually complementary variables, which changes as the incumbent changes. It diversifies itself, whereas a deterministic ranking of $|a_i|$ cannot.

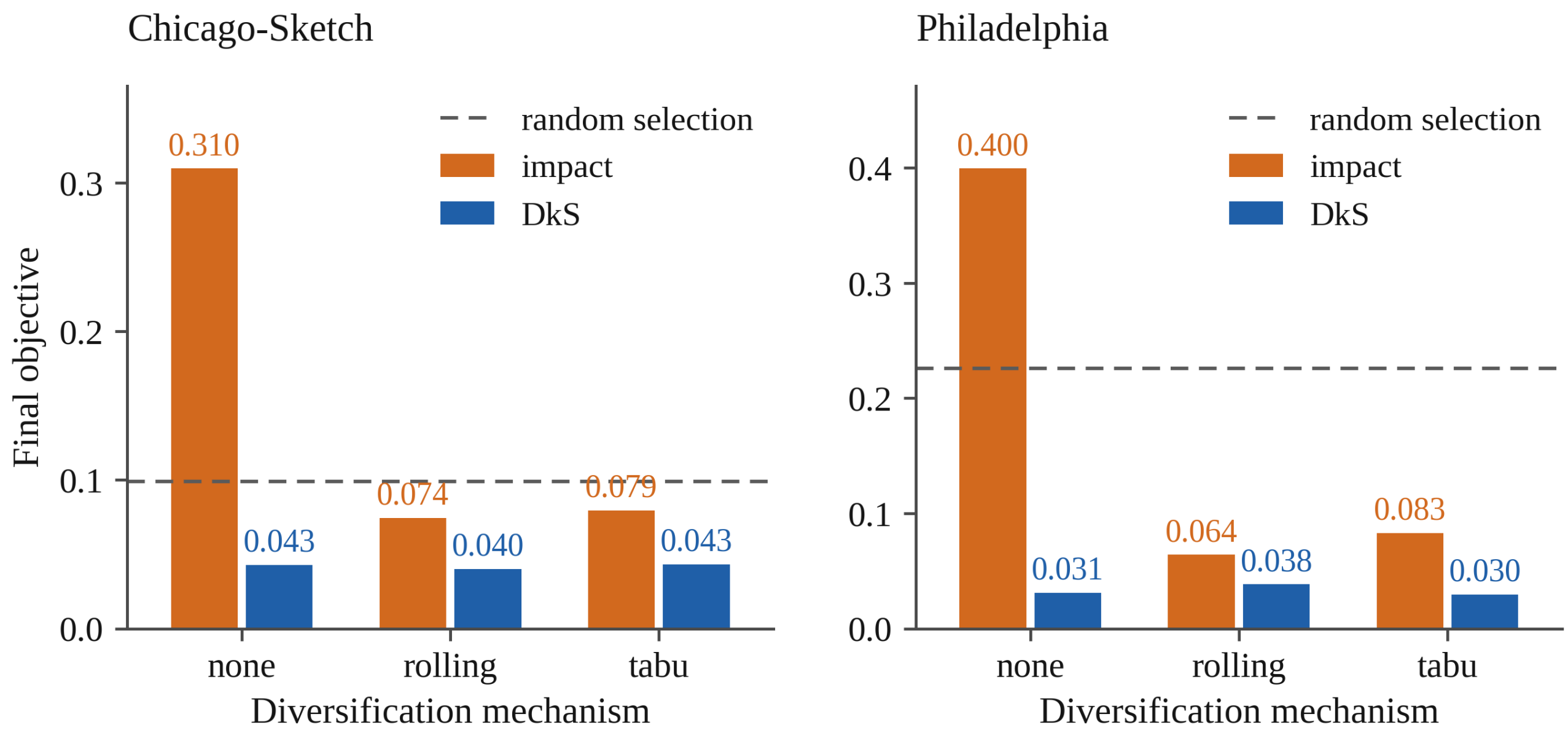

**Figure 9. Selection rule against diversification mechanism.** Full 2 × 3 ablation on both instances. The dashed line is the random-selection baseline. Impact collapses without diversification; DkS does not depend on it.

## 5.6 Where does the gain come from? A null ablation

All results so far use a classical sub-solver. The honest question is whether the improvements come from the selection rule or from sending sub-problems to a quantum computer. We answer it by holding the QUBO, the starting point, the selector and the number of rounds fixed, and sweeping the single quantity of interest: what fraction of the 20 rounds is solved on hardware. Table 8 gives the outcome.

**Table 8.** Configuration: Chicago-Sketch, Delaunay adjacency, $\gamma = 1.0$, $q = 16$, 20 rounds, DkS selector, single start from a fixed seed, identical selection trace in every column. The relative spread is exactly zero.

| Quantum share | 0% | 25% | 50% | 75% | 100% |
|---|---|---|---|---|---|
| Final objective | 0.097667 | 0.097667 | 0.097667 | 0.097667 | 0.097667 |

All five values are identical to every reported digit, and as Figure 11 shows, the five convergence trajectories coincide point for point. We therefore state the scope of the contribution explicitly: the improvement comes from the selection rule and the outer loop, and at this scale the quantum sub-solver is a substitutable backend.

The finding cuts both ways. Against the objection that the gain might be an artifact of the quantum step, the flat line is a direct negative answer. Against the objection that the quantum backend might not be solving correctly, 42 hardware calls all returned sub-problem optima identical to exhaustive enumeration, which is a direct positive answer. Figure 10 makes the same point on a single replayed trace, where the classical and quantum trajectories lie on top of one another round by round.

One limitation must be stated plainly. At $q = 16$ the classical sub-solver is exact, so the quantum backend cannot in principle exceed it. The flat line shows that the backend does not degrade the result, not that the two are of equal power. Testing the latter requires repeating the sweep where the classical solver is no longer exact. At $q = 64$, using our measured medians of 404 s QPU, roughly 300 s compilation and 819 s queueing per call, a single replicate costs about 2.2 hours of QPU and 10 hours of wall-clock time. Three replicates would cost 6.7 hours of QPU and roughly 25 hours of wall time.

As a by-product this answers the question of how many outer-loop rounds are needed. Convergence occurs at round 16. Rounds 13 and 14 produce no improvement but round 15 recovers a gain of 0.001538, direct evidence of the tabu penalty pushing selection toward unexplored variables, corroborating Section 5.5. A “stop after two rounds without improvement” criterion would therefore halt at round 13 and lose the later gain. We recommend five.

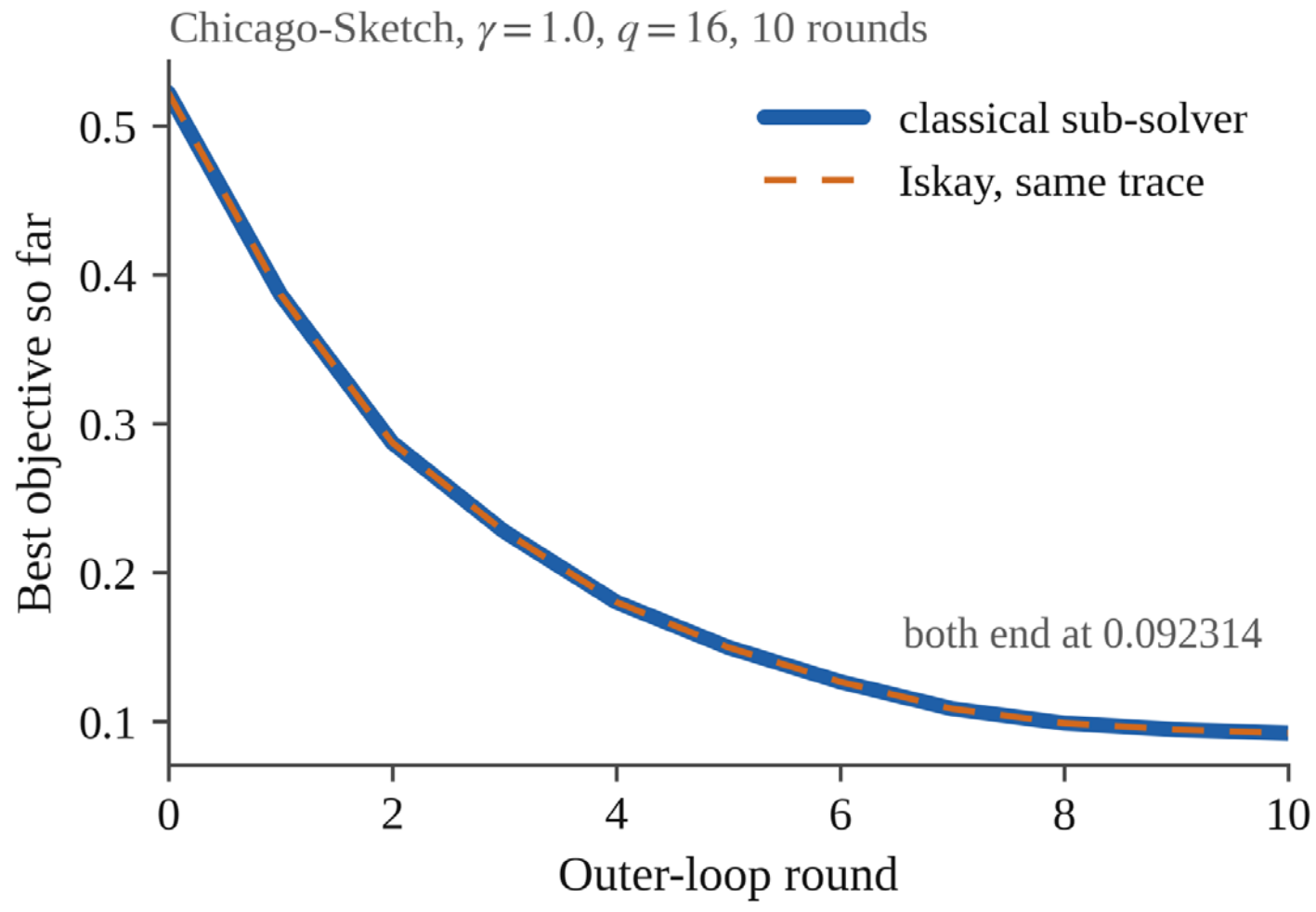

**Figure 10. The same selection trace under two sub-solvers.** Chicago-Sketch, Delaunay adjacency, $\gamma = 1.0$, $q = 16$, 10 rounds, single fixed start. The classical and Iskay trajectories coincide, with final objectives of 0.092314 and 0.092314. This replay uses 10 rounds, whereas the share sweep of Table 8 and Figure 11 uses 20, which is why the two final objectives differ.

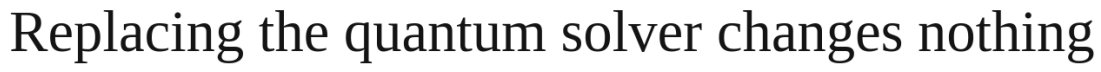

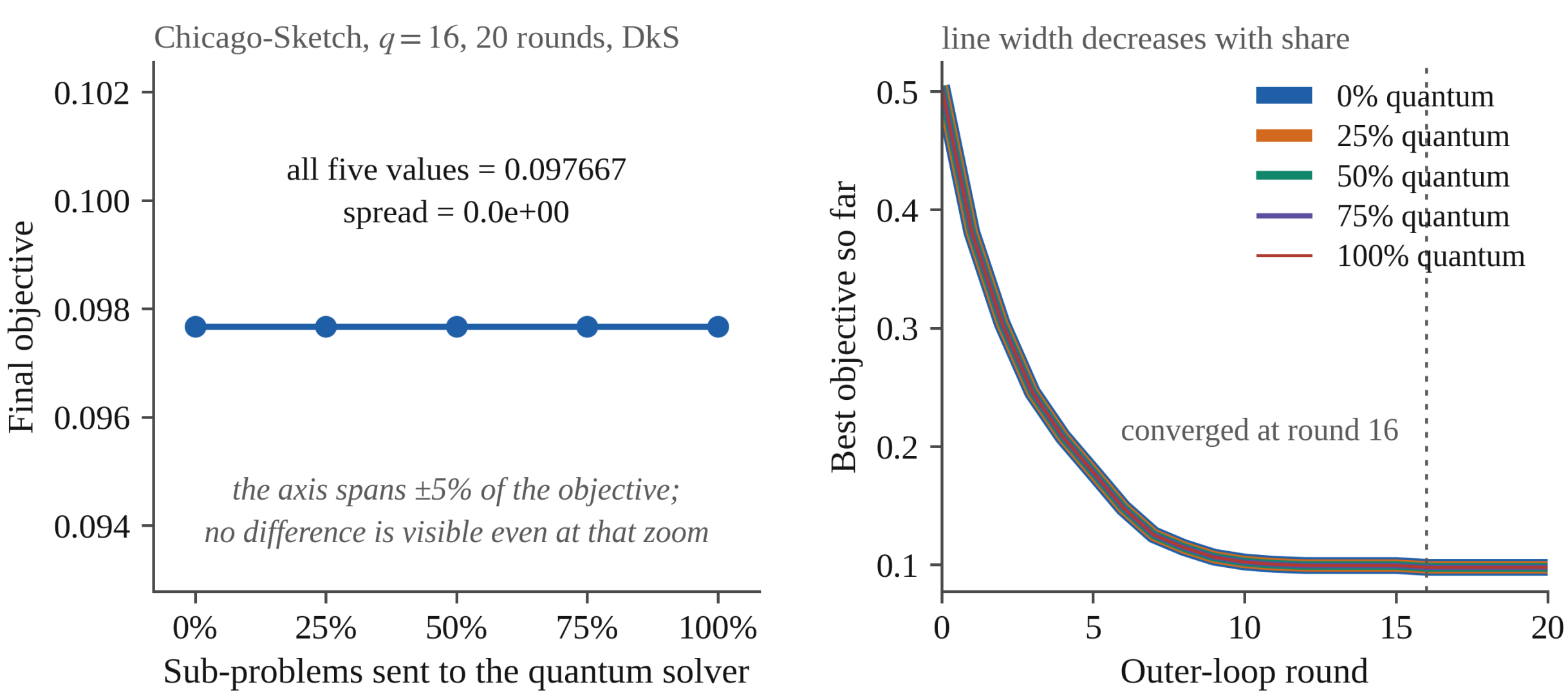

**Figure 11. Quantum-share sweep (null ablation).** Left: final objective against the fraction of sub-problems solved on hardware; the y-axis spans ±5% of the objective and no difference is visible even at that zoom. Right: the five convergence trajectories, drawn with decreasing line width so that each is visible; they overlap point for point.

### 5.7 Hardware limits: the wall is not the qubit count

All hardware runs use ibm_rensselaer as specified in Table 1. At every size where the device returned a result, the approximation ratio against a 60-second strong-tabu baseline was 1.0000, at $q$ = 16, 32, 64, 80 and 96, with three repetitions at each size and no failures. At $q$= 120, the full device width, 7,260 QUBO terms, all three jobs failed. This was not a timeout: hardware compilation returned an error after 282 seconds. Figure 12 shows the resulting step in job success rate, from 100 per cent at every size up to $q$ = 96 to zero at $q$ = 120.

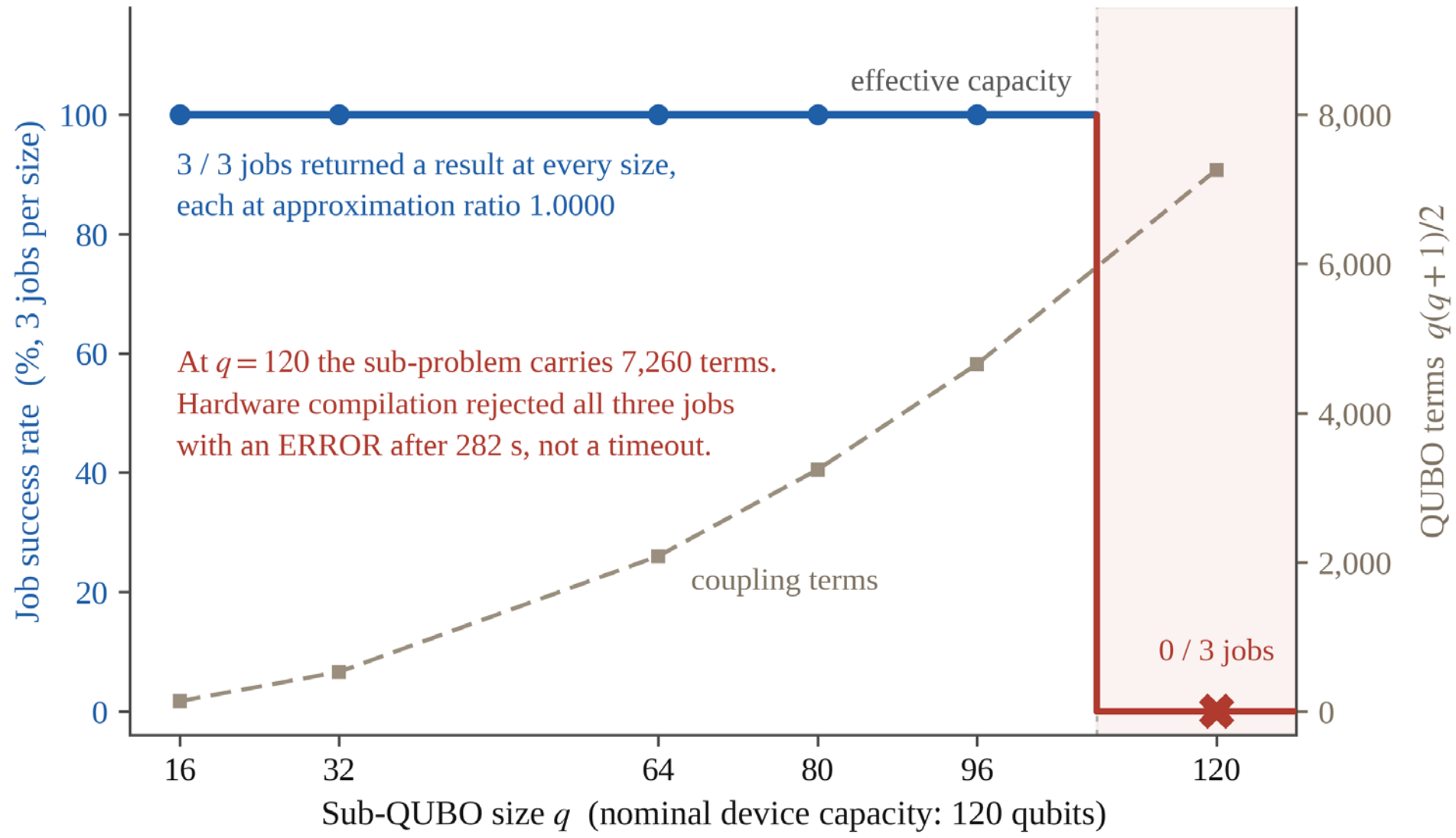

**Figure 12. Effective device capacity.** Job success rate against sub-QUBO size on ibm_rensselaer, three jobs per size, with the coupling-term count $q(q+1)/2$ on the right axis. The nominal capacity is 120 qubits, but the success rate falls from 100% to 0% between $q$ = 96 and $q$ = 120, where the term count reaches 7,260.

Whether that wall is caused by the qubit count or the term count can be separated by fixing $q = 120$ and sparsifying only the couplings, as Table 9 does.

**Table 9.** Configuration: Chicago-Sketch, γ = 1.0, q = 120, one sub-problem drawn from the first outer-loop round, three repetitions for the dense row and one for each sparsified row. All rows use 120 qubits and only the number of coupling terms varies. Compilation times are single measurements recovered from job status transitions, not medians.

| Couplings kept | Terms | Approx. ratio | Compile (s) | QPU (s) |
|---|---|---|---|---|
| 100% (dense) | 7260 | failed 3/3 | error at 282 | — |
| 70% | 5118 | 0.9981 | 1752 | 469 |
| 50% | 3690 | 0.9703 | 806 | 381 |
| 30% | 2262 | 0.8292 | 316 | 405 |
| 10% | 834 | 0.4895 | 56 | 356 |
| 3% | 335 | 0.1327 | 9 | 166 |

There is nothing wrong with $q = 120$ itself: at 70% sparsification the problem still occupies all 120 qubits and attains a ratio of 0.9981. The failure is triggered by the term count, not the variable count, and the boundary lies between 5118 terms, which succeeds, and 7260, which does not. That single comparison carries the claim, since both configurations use the same number of qubits and differ only in how many coupling terms the compiler is asked to handle.

The remaining rows trace the cost of buying compilability this way. Approximation quality falls monotonically as couplings are removed, from 0.9981 at 70 per cent to 0.1327 at 3 per cent, so sparsification is a trade rather than a free saving, and on this instance 70 per cent is the point at which the problem compiles while quality is essentially untouched. We do not attempt to separate how much of that decline is information lost by sparsification from how much is the device solving the reduced problem less well, because the claim above does not require it. The ratios are measured against a 60-second tabu baseline rather than against a certified optimum, so they are comparable with each other but should not be read as distances from optimality.

Compilation, not the QPU, is the bottleneck, and the two scale quite differently. QPU execution time grows slowly with term count, a median of 57 s at $q = 16$ rising to 377 s at $q = 96$, a factor of 6.6 across a 34-fold increase in terms. Hardware compilation grows almost quadratically over the same range. Fitting the five sparsified points gives

$$\textit{compile time} \approx 1.5 \times 10^{-4} \cdot \textit{terms}^{1.89} \ \textit{seconds}, \quad R^2 = 0.999$$

over a fifteenfold range in term count and a two-hundredfold range in time. Figure 13 plots both series, the quality curves on a common term-count axis on the left and the two timing curves on the right. The two curves cross at roughly 2,500 terms: below that the QPU dominates the cost of a call, above it compilation does. On the largest successful instance compilation took 1,752 seconds against 469 seconds on the QPU.

This fit also sharpens the interpretation of the dense failure. Extrapolated to 7,260 terms it predicts about 3,100 seconds of compilation, whereas the failed jobs returned an error after 282 seconds, roughly a tenth of the time the fit says the work would have required. The dense case therefore did not exhaust a time budget. It was rejected early, which is consistent with a resource or capacity limit inside the compiler rather than with slowness.

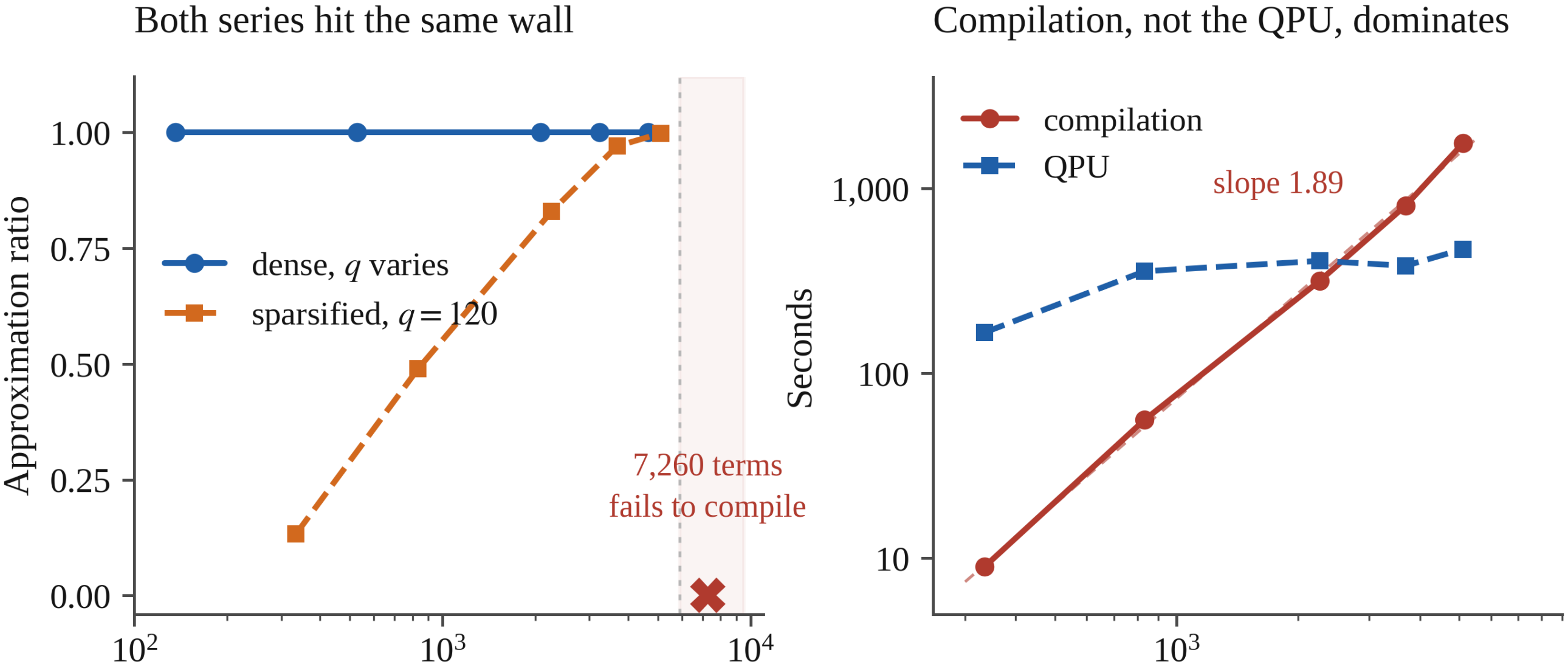

**Figure 13. The limit is coupling count, not qubit count.** Left: approximation ratio against the number of QUBO terms sent to the device, with the dense series ( $q$ varies) and the sparsified series ($q$ = 120 fixed) on a common axis; both reach the same wall. Right: compilation time and QPU time against term count, log-log.

Finally, the time budget. In the $q$ = 16 replay experiment, cumulative QPU time was 576 s against 10,158 s of wall-clock time, a QPU share of 5.7%. Queueing ranged from 103 s to 4,024 s, a 39-fold spread around a median of 819 s. Enan et al. (2024) report the same structure at a completely different problem scale: 2.99 s end-to-end of which 0.197 s is QPU access and 2.793 s is upload, queueing and download, a QPU share of 6.6%. Two independent setups, two orders of magnitude apart in problem size, agree that wall-clock time on current cloud quantum services is dominated by queueing. Any claim of quantum speedup measured in wall-clock time is therefore not reproducible, and we report QPU time from the provider's usage records throughout, listing compilation and queueing separately.

## 6 Practical guidance

Several conclusions here transfer to other sub-QUBO applications independently of the specific problem, and we state them plainly because they are the parts most likely to be reused.

Check the coupling distribution before expecting any structure-reading rule to help. The coefficient of variation of the off-diagonal entries in flip space is a one-line diagnostic. On our real instances it is 2.92 and 6.05 and the coupling-aware rule wins comfortably. On a synthetic instance with a value of 0.58 it is indistinguishable from random selection. A near-uniform coupling matrix carries no information for a selector to exploit, and no amount of algorithmic effort will manufacture any.

Do not rank by single-flip scores at a local optimum. The condition that makes impact indexing intuitive, large individual objective changes, is exactly the condition that makes it uninformative once the incumbent is 1-opt optimal, because all such changes are then costs. If a per-variable score must be used for reasons of simplicity, it needs an external diversification mechanism to function at all, and Section 5.5 measures that dependence at a factor of four to six.

Build stopping rules on the lower bound, not the upper bound. The upper bound continues to spike long after the incumbent has converged, because it sums every locally favorable term without regard to whether they are jointly attainable. The lower bound tracks the achievable improvement closely and costs the same to compute.

Do not stop after two rounds without improvement. In our runs the loop produced nothing at rounds 13 and 14 and then recovered a further gain at round 15, because the diversification penalty had pushed selection onto variables not yet examined. Five consecutive empty rounds is a safer criterion, and the cost of the extra rounds is small relative to the risk of stopping early.

Size sub-problems by coupling terms, not by qubits. A device advertised at n qubits will not necessarily accept an n-variable problem, because what the compiler must handle is the number of quadratic terms, and compilation time grows almost as the square of that count. Planning around the qubit count invites failures that appear, unhelpfully, as compilation errors rather than as capacity errors.

Report processing time from the provider's usage records, never wall-clock time. Queueing dominated wall-clock time in every experiment we ran, with a spread of 39 times between the fastest and slowest identical job, and the same pattern is reported independently at a problem scale two orders of magnitude smaller (Enan et al., 2024). A wall-clock comparison is not reproducible even by the same authors on the same hardware a day later.

## 7 Conclusion and future study

We have argued that, in a sub-QUBO decomposition, the choice of which variables to release is worth more than the choice of what solves the resulting sub-problem, and we have measured both. A selection objective derived from the second-order expansion of the QUBO at the incumbent outperforms the standard per-variable impact rule used by qbsolv (Booth et al., 2017) and carried into traffic zone partitioning in our own earlier work (Ke et al., 2026), by a margin that four times the device capacity does not close, and it comes with bounds that bracket the value of a selection before any solver is invoked. Replacing geometric adjacency with real road-network connectivity, in the spirit of partitioning studies that couple network structure with traffic attributes (Ma et al., 2023), widens that margin further, a small but concrete instance of domain knowledge paying for itself inside an algorithm rather than only in the problem formulation.

This finding depended on testing it in the right place. Our own early synthetic benchmark was degenerate: a balance term contributed almost all of the coupling mass, leaving the interaction structure nearly uniform, and on such an instance no structure-reading rule can beat random selection. Real origin-destination matrices are heavy-tailed, and once the benchmark reflected that, the difference became decisive. On the two real networks studied here, the evidence resolves into three claims, one of which is negative and which we regard as no less important than the other two.

The first claim is that selection substitutes for hardware. On a 1525-zone network, the proposed rule at a sub-problem size of 16 variables outperforms random selection at 64, so a fourfold increase in device capacity does not compensate for a weaker selection rule; the claim requires no quantum hardware, either to establish it or to act on it. The second claim follows from the first and sharpens it: domain structure widens the effect. Replacing the usual geometric adjacency between zone centroids with connectivity along the real road network, which changes 62% of the edges on the larger instance, widens the advantage of the coupling-aware rule by a factor of roughly 1.6 to 1.9, because the sparser adjacency makes the coupling structure more discriminative. Together, these two claims say that both the selection rule and the graph it reads matter more than the device it is eventually handed to.

The third claim qualifies the first two, and is the one we report as a deliberate negative result rather than leave for a reader to suspect: none of the observed gain is attributable to the quantum sub-solver. Sweeping the fraction of sub-problems sent to hardware from none to all, with the selection trace held identical, changes the final objective by nothing at all, five settings, one value, to every reported digit. This fixes the scope of the contribution precisely: the method is a classical selection rule for which a quantum device can serve as a backend, not a demonstration of quantum advantage.

Along the way, the hardware experiments produced a result of independent interest that reinforces the same theme. Running at the full width of a 120-qubit device fails, but not because of the qubit count: the same 120 variables succeed once the coupling terms are thinned, and the failure boundary tracks the number of terms rather than the number of variables. The binding cost is hardware compilation, which grows almost quadratically in the number of terms and, on our largest successful instance, took 1752 seconds against 469 seconds of quantum processing, a further illustration that structure, here in the compiled problem rather than in the selector, is what limits the method, not raw device size.

Two boundaries on these claims should be kept in view. All experiments are zone bipartition on two networks drawn from the Transportation Networks for Research repository, so the transferability asserted in Section 6 rests on the derivation, which does not depend on the problem family, rather than on cross-family evidence, which we have not gathered. And the quantum sub-solver, the Iskay Quantum Optimizer supplied by Kipu Quantum (2026), on the instances and sizes we could reach, contributed nothing that a classical solver did not. We prefer to state that explicitly. It bounds what the present work claims, and it identifies what would have to change for the question to be reopened: sub-problems large enough that exact classical solution is no longer available, which on current hardware means confronting the compilation cost that Section 5.7 measures rather than the qubit count that is usually quoted.

Three extensions follow naturally from where these boundaries sit. The first is to move beyond zone bipartition to other transportation optimization families, since the derivation itself is problem-family agnostic even though the empirical evidence here is not. The routing and dispatching problems already cast in QUBO or Ising form elsewhere in the transportation literature, cooperative platoon routing and dispatching (Azfar & Ke, 2026a), quantum-assisted vehicle routing (Azfar et al., 2025), public transit network design (Ke & Guo, 2026), and resilient network restoration (Udekwe et al., 2026), would each test whether the coupling-aware selector's advantage survives a change of problem family and not merely of instance.

The second follows directly from the compilation result above: since the binding hardware constraint is the number of coupling terms rather than the qubit count, hardware-efficient formulations that reduce term count directly, such as compressed adiabatic evolution (Azfar et al., 2026) and shallow, ramp-scheduled QAOA circuits (Azfar & Ke, 2026b), are natural companions to a selector that already keeps sub-problems small. Combining the two may push the useful sub-problem size past what compilation alone allows today.

The third is to test the selector against other emerging quantum optimization backends, since all hardware experiments reported here use a single gate-model backend (Kipu Quantum, 2026). Digitized counterdiabatic optimization has recently reported runtime advantages on higher-order binary optimization instances (Gomez Cadavid et al., 2025; Romero et al., 2025; Chandarana et al., 2025), and repeating the sweep of Section 5.6 on such a backend would test whether the negative finding, that backend choice does not move the final objective, holds only for the gate-model annealer configured here or is a more general property of the decomposition.

## 8 Acknowledgment

Q. Guo was partially supported by Natural Science Foundation under Grant No. 2400153, and the U.S. Department of Transportation (USDOT) Advancing Community-Centric Transportation Systems (ACTS) Center under Grant Number 693JK42550003. R. Ke is partially funded through the IBM-RPI Future of Computing Research Collaboration (FCRC).

## 9 Author Contributions

The authors confirm contribution to the paper as follows: Q. Guo: Data curation; Formal analysis; Investigation; Methodology; Visualization; Writing – original draft; Writing – review & editing. R. Ke: Conceptualization; Funding acquisition; Methodology; Resources; Validation; Writing – original draft; Writing – review & editing.